\documentclass[aps,prl,twocolumn,superscriptaddress,floatfix]{revtex4-2}
\usepackage[utf8]{inputenc}
\usepackage[english]{babel}
\usepackage{amsmath}
\usepackage{feynmf}
\usepackage{xcolor}
\usepackage{soul}
\usepackage{graphicx}
\usepackage{amsfonts}
\usepackage{blindtext}
\usepackage[colorlinks = true,
            linkcolor = blue,
            urlcolor  = blue,
            citecolor = blue,
            anchorcolor = blue]{hyperref}

\usepackage[normalem]{ulem}

\renewcommand{\vec}[1]{\mathbf{#1}}
\newcommand{\vk}{{\vec k}}
\newcommand{\vq}{{\vec q}}

\begin{abstract}
    In superconductors the Anderson-Higgs mechanism allows for the existence of a collective amplitude (Higgs) mode which can couple to eV-light mainly in a non-linear Raman-like process. 
The experimental non-equilibrium results on isotropic superconductors have been explained going beyond the BCS theory including the Higgs mode. Furthermore, in anisotropic \textit{d}-wave superconductors strong interaction effects with other modes are expected. 
Here we calculate the Raman contribution of the Higgs mode from a new perspective, including many-body Higgs oscillations effects and their consequences in conventional, spontaneous Raman spectroscopy. 
Our results suggest a significant contribution to the intensity of the $A_{1g}$ symmetry Raman spectrum in $d$-wave superconductors. 
In order to test our theory, we predict the presence of measurable characteristic oscillations in THz quench-optical probe time-dependent reflectivity experiments.
\end{abstract}

\begin{document}

\title{Calculation of an enhanced A$_{1g}$ symmetry mode induced by Higgs oscillations in the Raman spectrum of high-temperature cuprate superconductors}

\author{M.~Puviani}
\email{m.puviani@fkf.mpg.de}
\affiliation{Max Planck Institute for Solid State Research, 70569 Stuttgart, Germany}
\author{A.~Baum}
\affiliation{Walther Meissner Institut, Bayerische Akademie der Wissenschaften, 85748 Garching, Germany}
\author{S.~Ono}
\affiliation{Central Research Institute of Electric Power Industry, Yokosuka, 240-0196 Kanagawa, Japan}
\author{Y.~Ando}
\affiliation{Institute of Physics II, University of Cologne, 50937 Köln, Germany}
\author{R.~Hackl}
\affiliation{Walther Meissner Institut, Bayerische Akademie der Wissenschaften, 85748 Garching, Germany}
\author{D.~Manske}
\email{d.manske@fkf.mpg.de}
\affiliation{Max Planck Institute for Solid State Research, 70569 Stuttgart, Germany}

\date{\today}
\maketitle

\begin{figure}
    \centering
    \includegraphics[width=5cm]{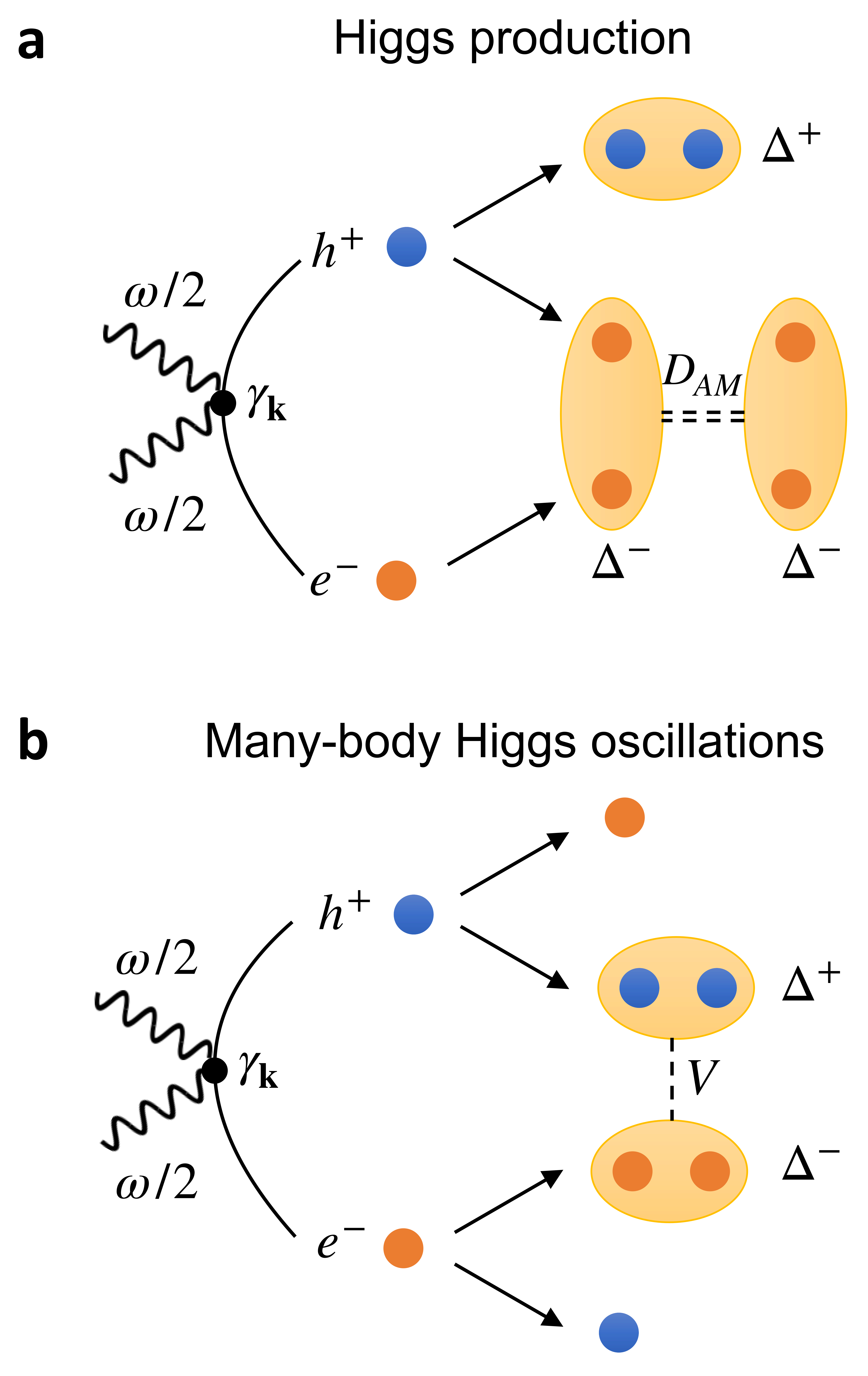}
    \caption{\textbf{Higgs production and many-body Higgs oscillations mechanisms}. 
    (\textbf{a}) Raman 2-photon interaction (vertex function $\gamma_\vk$) of light with the condensate leading to the creation of an electron-hole pair ($e^- - h^+$): when the hole (electron) undergoes Andreev scattering, it produces a Cooper pair and an electron (hole), and the latter forms a pair with the other electron (hole) thus producing a Higgs (\textbf{b}) Light creates an electron-hole pair: both particles undergo Andreev scattering and produce Cooper pairs which interact via many-body Higgs oscillations \cite{Supplement}.]}
    \label{model}
\end{figure}

\textit{Introduction.} Collective excitations of superconductors in non-equilibrium are a new emerging field. The most studied collective modes in superconductors, among others, are the amplitude (AM, so-called Higgs mode) and phase (Goldstone) modes \cite{PhysRev.130.439, Varma2002}. Yet, while the Higgs is a collective oscillation having a frequency of twice the superconducting order parameter $\Delta$, i.e. $\omega_{\text{Higgs}} = 2 \Delta$, the phase mode is shifted to the plasma frequency at higher energies due to the Anderson-Higgs mechanism \cite{PhysRev.130.439}. Since the Higgs mode does not carry a dipole moment, it is well known that a linear activation by light is almost impossible \cite{seibold2020harmonic, villegas2020anomalous}, such that the relevant process is the non-linear Raman effect. In ultra-fast non-equilibrium experiments the Higgs mode has been uncovered both in quench-probes \cite{PhysRevLett.111.057002} and in periodically driven setups \cite{Matsunaga1145, PhysRevB.96.020505, Chu2020}, respectively, where the coupling of light to the charges of the superconductor can be described by a quadratic  Raman-like process. In this context, the third harmonic generation has been a widely studied example of this non-linear optical effect \cite{PhysRevB.92.064508, PhysRevB.93.180507, PhysRevB.101.184519}.  \\
In the '90s conventional, steady-state Raman spectroscopy on superconductors turned out to be an effective experimental technique to provide evidence via polarization-dependence for $d$-wave Cooper pairing \cite{PhysRevLett.65.915, PhysRevB.43.12958, PhysRevB.55.12725, PhysRevB.29.4976, PhysRevB.41.8798, PhysRevLett.72.396}. In high-T$_{\text{c}}$ cuprates with $D_{4h}$ crystal symmetry the three most important Raman symmetries ($A_{1g}$, $B_{1g}$, $B_{2g}$) could be explained by a $d$-wave order parameter \cite{PhysRevLett.72.396, PhysRevB.51.16336, PhysRevB.54.15547,Kampf1997,RevModPhys.79.175}. However, in \textit{all} spectra of cuprates there is still an open question concerning the $A_{1g}$ peak intensity \cite{RevModPhys.79.175, PhysRevB.69.180501}. In conventional Raman scattering theory,  due to Coulomb screening of quasiparticles (QP) the response in $A_{1g}$ symmetry is much smaller than the experimental observations in cuprates \cite{RevModPhys.79.175, PhysRevB.41.8798}, which is of the same order or larger than the $B_{1g}$ peak \cite{PhysRevB.51.16336, PhysRevB.54.15547, PhysRevLett.72.396}.  This is the so-called long-standing $A_{1g}$ problem \cite{RevModPhys.79.175}. Over the past three decades, various attempts have been made to reconcile theory with experiments without arriving at conclusive results: these include, but are not limited to, the use of higher harmonics \cite{PhysRevB.51.16336, PhysRevB.54.15547}, magnon resonance \cite{PhysRevB.62.15204, PhysRevB.62.5525}, screening effects \cite{PhysRevB.56.R2940}, neutron magnetic resonance \cite{LETACON2006503} and the interaction of the AM with the $\eta-$mode, a spin-singlet excitation which can appear below $2 \Delta$ in $d$-wave superconductors \cite{PhysRevB.93.024515}.  However, none of these approaches could solve the $A_{1g}$ problem quantitatively, namely providing the correct intensity and peak position. \\
Recently, Puviani \textit{et al. }\cite{PhysRevB.101.220507} have shown that both equilibrium and non-equilibrium activation of the Higgs oscillations in superconductors correspond to the same physical non-linear Raman process. The conventional equilibrium Raman response of $s$-wave superconductors should be small for symmetry arguments, with the exception of the presence of competing orders, i.e. superconductivity and charge density wave \cite{PhysRevB.26.4883, PhysRevLett.122.127001}.
Eventually, recent terahertz pump-optical probe (TPOP) experiments have provided hints of the presence of the Higgs mode through transient reflectivity change in the \textit{d}-wave superconductor Bi2212 \cite{PhysRevLett.120.117001, PhysRevB.102.054510}. Despite this, no effective time-dependent oscillation detection, nor a consistent theoretical calculation have been provided for \textit{d}-wave superconductors to support this. \\
In this Letter, we go beyond the present theoretical background on the amplitude mode in superconductors,  including many-body effects of Higgs oscillations on quasiparticles: when light interacts with the charges of the condensate creating an electron-hole pair, these particles undergo the so-called Andreev scattering which leads to creation of pairs of electrons/holes which add to the existing condensate and eventually produce the Higgs (Fig. ~\ref{model}\textbf{a}). However, in addition to this production mechanism, which is generally subdominant, the quasiparticle excitation generated by light can lead to dominant many-body Higgs oscillations due to a mechanism of interaction between Cooper pairs (CPs) generated by Andreev scattering (Fig. ~\ref{model}\textbf{b}), which has been neglected so far. 
Theoretically, this is done through a systematic diagrammatic approach, adding vertex corrections in the amplitude channel alongside with the usual random phase approximation (RPA) summation for the AM propagator. This calculation gives rise to an effective hybridization of the amplitude and charge degrees of freedom, namely many-body effects of Higgs and quasiparticles, resulting now in a new strong contribution of the Higgs mode. This approach is able to show that there is a significant signature of the Higgs mode in the calculated Raman spectra of \textit{d}-wave superconductors, as suggested in Ref.s~\cite{PhysRevB.84.174522, PhysRevLett.110.140401}, generalizing the phenomenological coupling used by Zeyher and Greco in Ref. ~\cite{PhysRevLett.89.177004}. In our theory a qualitative and quantitative agreement with the Raman experimental data is possible. 
Simply speaking, similar to phonons which can be renormalized to polarons due to strong electron-phonon interactions, the Raman-active Higgs mode can be strongly renormalized due to quasiparticles to form a Cooper pair-polaron, resulting in a dressed Higgs mode with strong intensity in $A_{1g}$ polarization.  
Note that the conclusions of our work can be generalized to other clean systems (e.g.: SC+CDW, multi-band superconductors) where the Higgs peak is shifted from the quasiparticle continuum and the symmetry allows for a many-body response. \\
In order to further test our theory, we suggest that using an ultra-short THz pump in TPOP experiments our theory would explain the relative intensity of the transient reflectivity change measured for the $A_{1g}$ and $B_{1g}$ symmetries, and it predicts the possibility to detect the characteristic time-dependent oscillations due to the amplitude mode contribution in both of these symmetries. \\

\textit{Calculation of the amplitude mode.} For describing a \textit{d}-wave superconducting system, we consider a generalized version of the BCS theory,  namely the Hamiltonian: $H = H_{\text{el.}} + H_{\text{p}} + H_{\text{C}}$. It is the sum of the electronic tight-binding Hamiltonian $H_{\text{el.}}$,  a momentum-dependent attractive pairing interaction $H_{\text{p}}$ and a long-range Coulomb repulsive energy $H_{\text{C}}$, respectively. The pairing term responsible for the AM is given by
\begin{align}
    H_{\text{p}} = \sum_{\vk, \vk',\vq} V(\vk, \vk') \ \hat{P}_{\vk, \vq}^\dagger \ \hat{P}_{\vk',\vq} \,,
\end{align}
with the pair creation operator $\hat{P}^\dagger_{\vk,\vq} = \hat{c}^\dagger_{\vk+\vq, \uparrow} \hat{c}^\dagger_{- \vk, \downarrow}$, and the pairing interaction function $V(\vk, \vk')$  \cite{PhysRevB.92.064508, PhysRevB.93.180507}. For $d$-wave superconductors we can factorize the pairing interaction term $V(\vk, \vk') = -V f_\vk f_{\vk'}$, where $f_\vk = f^{d_{x^2-y^2}}_\vk$ is the $d_{x^2-y^2}$-wave form factor, V being the pairing strength derived from the self-consistent gap equation.  The Hamiltonian $H_{\text{el.}}+H_{\text{P}}$ can be treated at mean field level defining the order parameter $\Delta_\vk = \Delta_{\text{max}} f_\vk$, giving rise to a mean-field Hamiltonian for unconventional superconductors.  \\
The bare susceptibility response of the condensate coupling quadratically to light is given by
\begin{align} \label{susceptibility_gg}
    \chi_{\gamma \gamma} (\vq, \omega) = \sum_{\vk'} \gamma_{\vk'}^i \gamma_{\vk'}^s \dfrac{1}{\beta} \sum_{i \nu_m} \mathrm{Tr} \left[ G_{\vk',m}  \tau_3 G_{\vk',m}^+ \tau_3 \right] \,,
\end{align}
where $\tau_{i=1,2,3}$ are the Pauli matrices,  $\tau_0$ the identity matrix, $\gamma_\vk^{i,s}$ the interaction vertex for the incident/scattered light for a given symmetry, $G_{\vk', m} \equiv G(\vk', i \nu_m)$ and $G_{\vk',m}^+ \equiv G(\vk'+\vq, i \nu_m + i \omega)$ the matrices of the Matsubara Green's functions in Nambu-Gor'kov space. 
The coupling between the light and the amplitude mode occurs indirectly via the charge-amplitude susceptibility,  $\chi_{\gamma f} (\vq, \omega)$. 
Then, the AM propagator can be calculated with the RPA summation as in Ref.~\cite{PhysRevB.93.180507}
\begin{align}
	D_{\text{AM}}(\vq, \omega) = - \left[ 2/V + \chi_{f f} (\vq, \omega) \right]^{-1} \,. \label{Higgsprop}
\end{align}
Here we have introduced the susceptibility $\chi_{f f} (\vq, \omega)$ with vertices $f_\vk \tau_1 - f_\vk \tau_1$ in the amplitude channel.
This allows us to calculate the total Raman susceptibility including both quasiparticles (Eq. (\ref{susceptibility_gg})) and amplitude mode as
\begin{align} \label{RamanH}
    \tilde{\chi} (\omega) = \tilde{\chi}_{\gamma \gamma} (\omega) - \dfrac{\tilde{\chi}_{\gamma f}^2 (\omega) }{2/V + \tilde{\chi}_{f f} (\omega)} \,,
\end{align}
where we used the susceptibilities $\tilde{\chi}$ which include charge fluctuations due to the long-range Coulomb interaction between quasiparticles \cite{Supplement}. We want to stress here that for a given value of the order parameter $\Delta_{\text{max}}$ we calculate self-consistently within the gap equation the value of pairing strength V. \\
Note that here we have considered only the diamagnetic susceptibility, since the paramagnetic contributions are vanishing in clean superconductors and in visible light experiments, as well as coupling to the spin channel (resonance peak) is negligible \cite{Supplement}. \\

\textit{Many-body Higgs oscillations.} As mentioned earlier, we now want to go beyond the RPA summation for the amplitude mode propagator, including many-body dynamic Higgs interactions between Cooper pairs generated by electrons/holes via Andreev scattering (Fig.~\ref{model} \textbf{b}), mixing the amplitude and charge channels.  To do that, we defined a dressed light-condensate interaction vertex through the ladder diagram summation
\begin{align} \label{Gamma2}
	\Gamma (\vk, \omega) = \gamma_\vk \tau_3 - \dfrac{V}{2} f_\vk \sum_{\vk'} f_{\vk'} \dfrac{1}{\beta} \sum_{i \nu_m} & \big[ \tau_1 G_{\vk',m} \Gamma (\vk', \omega) \notag \\
	& \times G^+_{\vk', m} \tau_1 \big] \,.
\end{align}
Analogously, for the $f_\vk \tau_1$ vertex in the amplitude channel we derived the self-consistent sum
\begin{align} \label{Lambda2}
	\Lambda (\vk, \omega) = \tau_1 - \dfrac{V}{2} f_\vk \sum_{\vk'} f_{\vk'} \dfrac{1}{\beta} \sum_{i \nu_m} \big[ & \tau_1 G_{\vk',m} \Lambda (\vk', \omega) \notag \\
	& \times G^+_{\vk', m} \tau_1 \big] \,.
\end{align}
Substituting to the bare vertices the dressed ones including the many-body effects of the Higgs mode,  the Raman response in Eq. (\ref{RamanH}) can be replaced with the full many-body expression
\begin{align} \label{susceptibility_tot}
    \tilde{\chi}_{\text{full}} (\omega) = \tilde{\chi}_{\Gamma \Gamma} (\omega) - \dfrac{\tilde{\chi}^2_{\Gamma \Lambda} (\omega) }{2/V + \tilde{\chi}_{\Lambda \Lambda} (\omega)} \,,
\end{align}
where the susceptibilities $\tilde{\chi}$ include charge fluctuations \cite{PhysRevB.41.8798, PhysRevB.93.180507}. \\

\textit{Steady-state Raman spectroscopy.} 
The intensity measured in Raman experiments is given by the photon scattering differential cross section, which is proportional to the imaginary part of the susceptibility, namely $\partial^2 \sigma / \partial \omega \partial \Omega \propto - \tilde{\chi}'' (\vq, \omega)$. 
\begin{figure*}
    \centering
    \includegraphics[width=18cm]{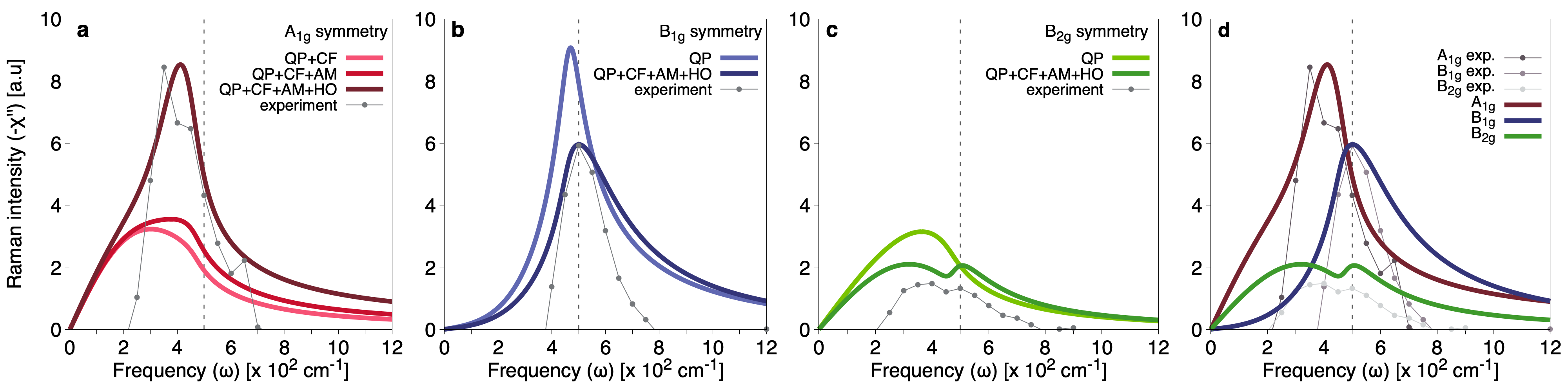}
    \caption{\textbf{Raman spectra of Bi2212}. Raman spectra contributions of various symmetries of a $D_{4h}$ crystal: comparison between different theoretical contributions and experimental results (represented by dots: the lines connecting them are a guide for the eyes) for Bi2212 \cite{Supplement}. The dashed vertical lines are placed at $\omega = 2 \Delta_{\text{max}}$, tuned with the $B_{1g}$ peak position. (\textbf{a}) $A_{1g}$ symmetry: the light red line is the quasiparticle (QP)+charge fluctuations (CF) contribution,  the intermediate one includes the amplitude mode (AM), the darker line includes also the many-body effects of the Higgs oscillations (HO).  (\textbf{b}) $B_{1g}$ spectrum intensity due to QP (light blue) and adding the CF,  AM and HO contributions (the QP+CF+AM plot would coincide with QP).   (\textbf{c}) $B_{2g}$ Raman spectrum response of QP (light green) and including CF and the full Higgs contribution (the QP+CF+AM plot would coincide with QP).  (\textbf{d}) The responses for all the symmetries are placed together.}
    \label{Raman_fig}
\end{figure*}
In order to solve the set of equations for the dressed vertices and the susceptibilities, we considered the lowest terms in a Fermi-surface harmonic expansion of the Raman vertices, restricting ourselves to $\vq=0$ for the light and to the clean limit of superconductor, in agreement with previous studies \cite{PhysRevB.51.16336, PhysRevB.54.15547}. Using the same optimized parameters fixed for all the Raman symmetries investigated, we additionally checked systematics in order to confirm the stability of the results \cite{Supplement}. In Fig. ~\ref{Raman_fig}\textbf{a} we show the effect of the bare AM and the many-body dressing contribution in the Raman spectra for different symmetries of Bi2212 (a \textit{d}-wave superconductor with $2 \Delta_{\text{max}} \sim 550 \ \text{cm}^{-1}$): on the one hand, for the $A_{1g}$ response the dressed AM calculated within the RPA approximation using Eq. (\ref{RamanH}) is negligible, while the full many-body Higgs oscillations result into a contribution which is more intense and peaked. On the other hand, for the $B_{1g}$ and $B_{2g}$ symmetries the charge fluctuations and Higgs oscillations screen the Raman response, reducing its intensity.  In particular, the $B_{1g}$ spectrum becomes broader in frequency and the peak gets shifted to higher frequencies, while the $B_{2g}$ acquires a shoulder peaked at around $\omega \sim 2 \Delta_{\text{max}}$. \\
These many-body Higgs effects describe the experimental Raman on Bi2212 in a much improved fashion, thus contributing to the long-standing problem regarding the $A_{1g}$ peak intensity \cite{PhysRevB.51.16336, PhysRevB.54.15547}. As we have concluded before, the $A_{1g}$ response provided by QPs is much lower than the experimental one, and the bare AM contribution is negligible. However, considering many-body effects of the Higgs oscillations, the $B_{1g}$(Fig. ~\ref{Raman_fig}\textbf{b}) and $B_{2g}$ (Fig. ~\ref{Raman_fig}\textbf{c}) spectra get screened, while the $A_{1g}$ response becomes enhanced and peakes now at a frequency $\omega \sim 400 \ \text{cm}^{-1}$, in quite good agreement with the experimental result.  Note that a simultaneous and quantitative understanding of all relevant polarizations is indeed possible employing  the \textit{same} set of many-body parameters \cite{Supplement} as shown in Fig. ~\ref{Raman_fig}\textbf{d}. 
This contributes to explain the long-standing $A_{1g}$ problem concerning the peak intensity in optimally doped Bi2212 as formulated in the literature \cite{PhysRevLett.72.396,PhysRevB.51.16336,RevModPhys.79.175}.  \\

\textit{Ultra-fast Raman-like optics.} As discussed earlier, the same Higgs oscillations should be seen in non-equilibrium in a time-dependent experiment. While, in principle, it is possible to measure Fig. ~\ref{Raman_fig}\textbf{d} time-dependently in an ultra-fast experiment, it is much easier to detect the Higgs oscillations in the transient reflectivity change $\Delta \text{R} (t) / \text{R}$. 
It has been widely demonstrated that the coupling of the light to the collective AM can happen only in a nonlinear regime with a Raman-like interaction, thus new evidences should be looked for in nonlinear optical effects, such as the Kerr signal in pump-probe experiments. To this extent, it is useful to look at the time-dependent reflectivity change in non-equilibrium superconductors, which is related to the equilibrium Raman susceptibility aforementioned and easily accessible in experiments. The transient reflectivity change due to a pump electric pulse $E_{\text{pump}}(t)$ is given by \cite{He2014}
\begin{align} \label{DeltaR}
    \dfrac{\Delta \text{R}}{\text{R}}(t) \propto \int_{- \infty}^{t} \chi'^{(3)}(t-t') \ E^2_{\text{pump}}(t') \ \text{d}t' \,,
\end{align}
where $\chi'^{(3)} (t-t')$ is the real part of the third order susceptibility in real time, $t$ being the time at which the reflectivity change is detected, $t'<t$ the previous time at which the pump pulse interacts with the condensate.
\begin{figure}
    \centering
    \includegraphics[width=7.5cm]{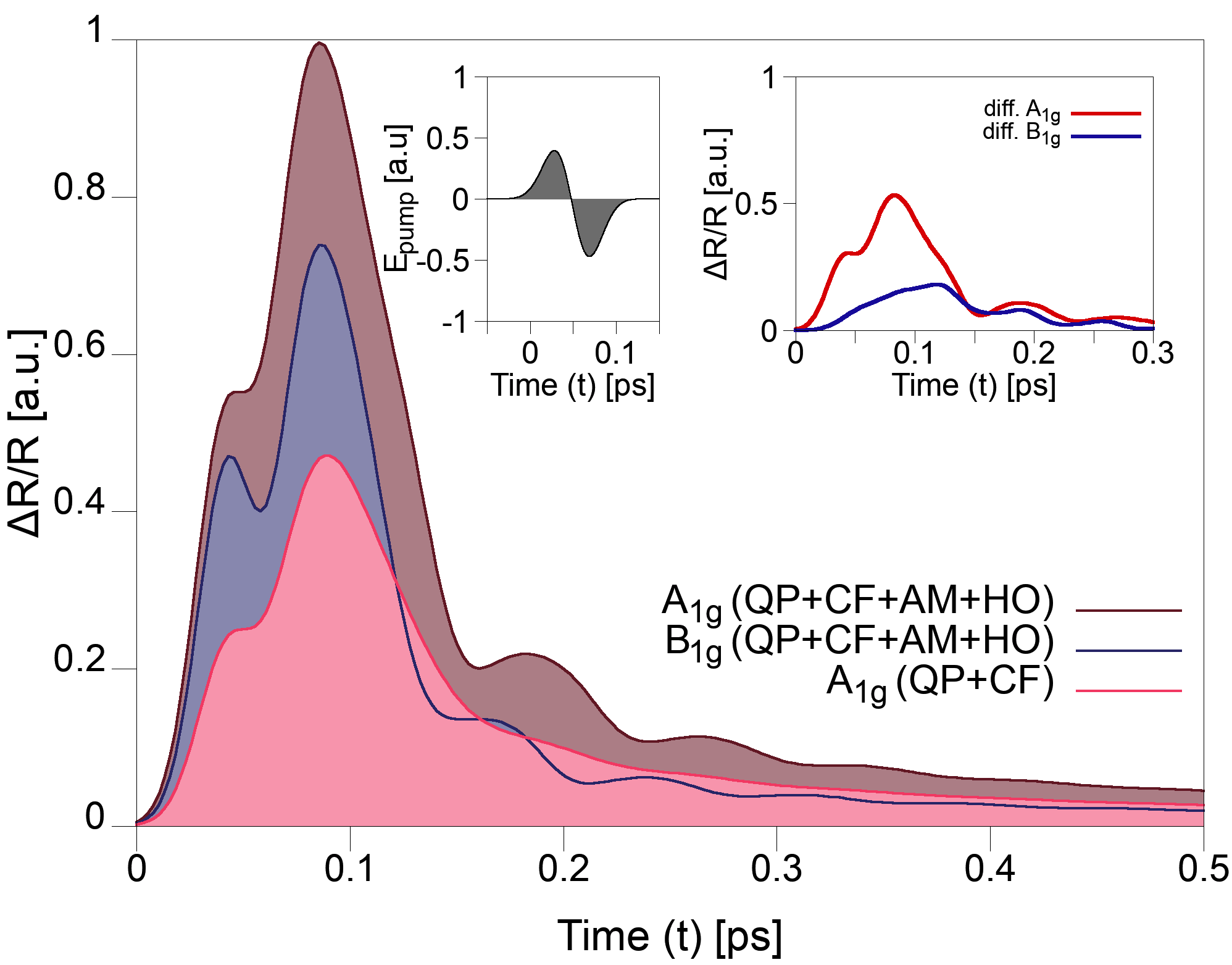}
    \caption{\textbf{Transient reflectivity}. Normalized transient reflectivity change $\Delta \text{R} (t)/\text{R}$ after a quench, obtained applying an electric pump pulse ($E_{\text{pump}}$, shown in the left inset) with duration $\Delta t_{\text{pump}} \sim 0.1 \ \text{ps}$ and frequency $\omega_{\text{pump}} = 5.57 \ \text{THz}$. The dark red and dark blue curves represent the normalized transient reflectivity of the $A_{1g}$ and $B_{1g}$ symmetries, respectively, calculated including the full many-body Higgs contribution. The light red curve, instead, shows only the QP and CF response in the $A_{1g}$ symmetry. The top right inset shows the differential $A_{1g}$ (red line) and $B_{1g}$ (blue line) signal due to the Higgs contribution.}
    \label{DeltaR_fig}
\end{figure}
Recently, Katsumi et al. \cite{PhysRevLett.120.117001, PhysRevB.102.054510} demonstrated that the relative intensity between the transient reflectivity change projected onto the symmetries $A_{1g}$ and $B_{1g}$ in a TPOP experiment on Bi2212 cannot be explained only with the quasiparticle density fluctuations contribution, deducing that the Higgs should play an important role.  However, since they pumped the system in the fully adiabatic regime ($\omega_{\text{pump}} \ll \Delta_{\text{max}}, \Delta t_{pump} \sim 4 \ \text{ps}$), they were not able to probe any Higgs oscillation in the time-dependent response.\\
We performed the calculation of the theoretically expected time-dependent transient reflectivity change on a \textit{d}-wave superconductor, using Eq. (\ref{DeltaR}) together with the third-order susceptibility containing the full many-body Higgs contribution according to Eq. (\ref{susceptibility_tot}). Considering the same pump regime of the experiments, we have been able to get a good agreement of the relative intensities for the $A_{1g}$ and $B_{1g}$ symmetries \cite{Supplement}. In addition to this, we repeated the numerical calculation with an ultra-short THz pump (duration $\Delta t_{\text{pump}} \sim 0.1$ ps, frequency $\omega_{\text{pump}} = 5.57$ THz $\lesssim \Delta_{max}$) in order to simulate a quench experiment: the result is shown in Fig. ~\ref{DeltaR_fig}. Including the full Higgs contribution to the third-order susceptibility, oscillations with a frequency $\omega = \omega_{\text{Higgs}} = 12.5 \ \text{THz} \lesssim 2 \Delta_{\text{max}}$ appear in the decaying region of $\Delta \text{R} (t)/ \text{R}$, while a lower intensity and no oscillations are present considering only the QPs and charge fluctuations contributions. Performing such a time-resolved experiment would provide a clear evidence of the presence of the Higgs mode and the effects of many-body contributions in \textit{d}-wave superconductors: indeed, they are responsible for enhancing the $A_{1g}$ symmetry response and, at the same time, for the screening effect on the $B_{1g}$. \\

\textit{Conclusion and outlook.} For the first time, we provide calculations of Higgs oscillations for \textit{d}-wave superconductors including many-body contributions originating from the interaction with quasiparticles (e.g. \textit{broken} Cooper pairs), resulting in a renormalized, dressed Higgs mode. We treat the full vertex corrections within a diagrammatic approach, which is able to describe both non-equilibrium and conventional steady-state Raman experiments, respectively. In particular, we are able to provide a contribution to the long-standing $A_{1g}$ problem in Raman spectra of $d$-wave superconductors \cite{PhysRevB.51.16336, PhysRevB.54.15547} in Fig. ~\ref{Raman_fig}\textbf{d}. Our conclusions are not based on any specific scenario for cuprates.
Furthermore, our results are also unchanged if we take into account that Higgs oscillations can be dressed by phonons as well. However, these contributions to the Raman spectra are minor \cite{Supplement}. Furthermore, we explain the non-equilibrium response and the correct magnitude of the transient reflectivity change in recent THz pump-optical probe experiments \cite{PhysRevLett.120.117001}. Our theory can simply be tested by a smoking gun experiment: we propose an experimental set-up using ultra-short THz pump pulse in a non-adiabatic regime to measure the reflectivity in \textit{d}-wave superconductors, in order to quench the condensate and thus directly observe the time-dependent oscillations characteristics of the Higgs mode.  \\
It is expected that the resonance peak obtained with inelastic neutron scattering can play a role in the $A_{1g}$ problem  because of the energy matching with the $A_{1g}$ Raman response. However, it is interesting to note that there is no linear coupling of the spin channel to the Higgs mode or to the Raman vertex: as shown by Venturini et al. in \cite{PhysRevB.62.15204}, only including spin fluctuations at higher order it is possible to match the energy position of the $A_{1g}$ peak, but without substantially affecting its intensity. \\
We also point out that so far it has been possible to clearly detect the Higgs contribution only in the presence of competing orders such as in $2H-\text{NbSe}_2$, namely superconductivity and charge density wave. The latter provides the excited phonons which couple to the superconducting AM \cite{PhysRevB.90.224515}. This mechanism allows to shift the Higgs $A_{1g}$ peak to energies lower than $2 \Delta$, becoming thus unambiguously detectable \cite{PhysRevLett.45.660,PhysRevB.89.060503, Pekker-Varma, PhysRevB.90.224515}. 
Similarly, a Higgs response has been reported in $E_{2g}$ Raman response of pressure-induced $2H-\text{TaS}_2$ \cite{PhysRevLett.122.127001}, a transition metal dichalcogenide with cohexisting SC and CDW.
Our findings provide a new route for calculations of many-body effects of the Higgs mode while interacting with various other modes in superconductors \cite{PhysRevResearch.2.023413, PhysRevLett.45.660}. We believe that an extension of our model to these systems can provide further insights and guide future experiments on unconventional superconductors and systems with cohexisting phases.

\paragraph{Acknowledgements} 
Fruitful discussions with P.M. Bonetti, D. Einzel, Y. Gallais, A. Greco, R. Heid, S. Kaiser, L. Schwarz,  R. Shimano, D. Vilardi, R. Zeyher are thankfully acknowledged. We thank the Max Planck-UBC-UTokyo Center for Quantum Materials for fruitful collaborations and financial support.

\bibliography{Bibliography.bib}

\end{document}


\title{Supplemental Material for \\ Calculation of an enhanced A$_{1g}$ symmetry mode induced by Higgs oscillations in the Raman spectrum of high-temperature cuprate superconductors}

\author{M.~Puviani}
\email{m.puviani@fkf.mpg.de}
\affiliation{Max Planck Institute for Solid State Research, 70569 Stuttgart, Germany}
\author{A.~Baum}
\affiliation{Walther Meissner Institut, Bayerische Akademie der Wissenschaften, 85748 Garching, Germany}
\author{S.~Ono}
\affiliation{Central Research Institute of Electric Power Industry, Yokosuka, 240-0196 Kanagawa, Japan}
\author{Y.~Ando}
\affiliation{Institute of Physics II, University of Cologne, 50937 Köln, Germany}
\author{R.~Hackl}
\affiliation{Walther Meissner Institut, Bayerische Akademie der Wissenschaften, 85748 Garching, Germany}
\author{D.~Manske}
\email{d.manske@fkf.mpg.de}
\affiliation{Max Planck Institute for Solid State Research, 70569 Stuttgart, Germany}

\date{\today}
\maketitle

\section{I. Raman response calculation}

\subsection{A. Notation and definitions}
We provide here some definitions and short notation that is used throughout the text.\\
First, we define averages $\langle g(\vk) \rangle$ in terms of the Tsuneto function:
\begin{align}
        \langle g(\vk) \rangle & = \sum_\vk g(\vk) \dfrac{\tanh{(\beta E_\vk/2)}}{2 E_\vk^2} \left( \dfrac{1}{2 E_\vk + i \omega_n} + \dfrac{1}{2 E_\vk - i \omega_n} \right) \notag \\
        & = \sum_\vk g(\vk) \dfrac{\tanh{(\beta E_\vk/2)}}{E_\vk} \dfrac{2}{4 E^2_\vk - (i \omega_n)^2} \,,
        \label{average}
\end{align}
where $i \omega_n$ is the Matsubara imaginary frequency, $\eps$ the electronic band structure, $\E = \sqrt{\eps^2 + \D^2}$ the quasiparticles energy, while the gap function is $\D = \Delta_{\text{max}} f_{\vk}$.\\
Then, we can introduce the Nambu-Gor'kov spinor $\Psi^\dagger_\vk = (\hat{c}^\dagger_{\vk, \uparrow} \quad \hat{c}_{- \vk, \downarrow})$, and distinguish the interaction terms $\Psi^\dagger_\vk \tau_i \Psi_\vk$, where $\tau_i$ are the Pauli matrices: $\Psi^\dagger_\vk \tau_0 \Psi_\vk = \hat{c}^\dagger_{\vk, \uparrow} \hat{c}_{\vk, \uparrow} + \hat{c}_{- \vk, \downarrow} \hat{c}^\dagger_{- \vk, \downarrow}$ is the spin/current channel, $\Psi^\dagger_\vk \tau_1 \Psi_\vk = \hat{c}^\dagger_{\vk, \uparrow} \hat{c}^\dagger_{- \vk, \downarrow} + \hat{c}_{- \vk, \downarrow} \hat{c}_{\vk, \uparrow}$ the amplitude channel, $\Psi^\dagger_\vk \tau_2 \Psi_\vk = -i (\hat{c}^\dagger_{\vk, \uparrow} \hat{c}^\dagger_{- \vk, \downarrow} - \hat{c}_{- \vk, \downarrow} \hat{c}_{\vk, \uparrow})$ the phase channel and eventually the charge channel is given by $\Psi^\dagger_\vk \tau_3 \Psi_\vk = \hat{c}^\dagger_{\vk, \uparrow} \hat{c}_{\vk, \uparrow} - \hat{c}_{- \vk, \downarrow} \hat{c}^\dagger_{- \vk, \downarrow}$. \\
Thus, we write the bare BCS Green's functions in Nambu-Gor'kov space as
\begin{equation}
\begin{split}
    G (\vk, i \nu_m) & = \dfrac{i \nu_m \tau_0 + \varepsilon_{\vk} \tau_3 + \Delta_{\vk} \tau_1}{(i \nu_m)^2 - E_{\vk}^2} \\
    & = \dfrac{1}{(i \nu_m)^2 - E_{\vk}^2} \cdot 
    \begin{pmatrix}
        i \nu_m + \varepsilon_{\vk} & \Delta_\vk \\
        \Delta_{\vk} & i \nu_m - \varepsilon_{\vk}
    \end{pmatrix}  \,. 
\end{split}
\end{equation}
We can usefully rewrite it in its spectral form \cite{bruus_many-body_2004}
\begin{eqnarray}
    G (\vk, i \nu_m) = - \dfrac{1}{\pi} \int_{- \infty}^{+ \infty} \text{d} \omega \dfrac{G''(\vk, \omega + i \delta )}{i \nu_m - \omega} \,, \label{spectral}
\end{eqnarray}
where $G''(\vk, \omega + i \delta )$ is the imaginary part of the Green's function. In order to shorten the notation , we define $G_{\vk, m} = G(\vk, i \nu_m)$ and $G^+_{\vk, m} = G(\vk + \vq, i \nu_m + i \omega_n)$. The function in real frequency space is obtained with the analytic continuation $i \omega_n \rightarrow \omega + i \delta$.\\
Finally, we define the susceptibilities according to the interaction vertices, so that $\chi_{\tau_i \tau_j}$ are the polarization bubbles with $\tau_i - \tau_j$ interaction vertices, while $\chi_{\gamma \tau_i}$ is the $\gamma_\vk \tau_3 - \tau_i$ bubble and $\chi_{\gamma f}$ is the susceptibility with vertices $\gamma_\vk \tau_3 - f_\vk \tau_1$. Analogously we can write all the other susceptibility amplitudes.

\subsection{B. Bare Raman response}
For describing the \textit{d}-wave superconducting phase, we consider a model Hamiltonian which can be split into three components, namely $H = H_{el.} + H_{p} + H_{C}$, which is the sum of the electronic tight-binding term, attractive pairing interaction and long-range Coulomb repulsion, respectively. 
The tight-binding term is simply 
\begin{align}
    H_{\text{el.}} = \sum_{\vk, \sigma} \varepsilon_{\vk} \hat{c}^\dagger_{\vk, \sigma} \hat{c}_{\vk, \sigma} \,, \label{Hband}
\end{align} 
where $\varepsilon_\vk$ is the band dispersion; while the Coulomb interaction is given by 
\begin{align}
    H_{\text{C}} = \dfrac{1}{2} \sum_\vk \hat{c}_{\vk, \uparrow}^\dagger \hat{c}_{\vk, \uparrow} \ \dfrac{4 \pi e^2}{\vk \cdot \vk} \ \hat{c}^\dagger_{-\vk, \downarrow} \hat{c}_{- \vk, \downarrow} \,.
\end{align}
The pairing term responsible for the amplitude mode is given by
\begin{align}
    H_{\text{p}} = \sum_{\vk, \vk'} V(\vk, \vk') \ \hat{P}_{\vk, \vq}^\dagger \ \hat{P}_{\vk',\vq} \,, \label{Hpair}
\end{align}
with the pair creation operator $\hat{P}^\dagger_{\vk,\vq} = \hat{c}^\dagger_{\vk+\vq, \uparrow} \hat{c}^\dagger_{- \vk, \downarrow}$, and $V(\vk, \vk') = - V f_\vk f_{\vk'}$ the factorized pairing interaction function \cite{PhysRevB.92.064508, PhysRevB.93.180507} calculated as
\begin{align}
    V = \left[\sum_\vk \dfrac{f^2_\vk}{2 E_\vk} \tanh{(\beta E_\vk/2)} \right]^{-1} \,.
\end{align}
A simple choice for the tight-binding band structure to be used in (\ref{Hband}) which contains the basic physics of 2D-like tetragonal systems is given by the following:
\begin{eqnarray}
    \varepsilon_{\vk} = - 2 t \left[ \cos{k_x} + \cos{k_y} - 2 B \cos{k_x} \cos{k_y} + \mu / 2 \right] \,,  
    \label{dispersion}
\end{eqnarray} 
where $t$ is the nearest-neighbor hopping integral, $2 B = t'/t$ the ratio between the nearest-neighbor and the next-nearest-neighbor hopping, $\mu t$ the chemical potential. The \textit{d}-wave order parameter is $\Delta_{\vk} = \Delta_{\text{max}} f_\vk $ and the corresponding form factor results to be $f_\vk = (\cos{k_x} - \cos{k_y})/2$. Within the effective mass approximation, the Raman vertex can be written in terms of the curvature of the energy band according to the expression $\gamma_\vk = \sum_\vk \mathbf{e}_i \  \partial^2_{ij} \varepsilon_\vk \  \mathbf{e}_j$, with $\partial^2_{ij} = \partial^2/\partial k_i \partial k_j$, $\mathbf{e}_{i,j}$ being the polarization vectors of the incident and scattered light, respectively \cite{RevModPhys.79.175, PhysRevB.51.16336}. The derivatives of the band dispersion (\ref{dispersion}) with respect to $k_x$ and $k_y$ are 
\begin{align}
    & \partial_x \varepsilon_{\vk} = 2 t (\sin{k_x} - 2 B \sin{k_x} \cos{k_y}) \,, 
    \qquad \partial_y \varepsilon_{\vk} = 2 t (\sin{k_y} - 2 B \cos{k_x} \sin{k_y}) \,, \notag
\end{align}
and the second derivatives
\begin{align}
    & \partial^2_{xx} \varepsilon_{\vk} = 2 t (\cos{k_x} - 2 B \cos{k_x} \cos{k_y}) \,,
    \qquad \partial^2_{yy} \varepsilon_{\vk} = 2 t (\cos{k_y} - 2 B \cos{k_x} \cos{k_y}) \,,
    \qquad \partial^2_{xy} \varepsilon_{\vk} = 4 t B \sin{k_x} \sin{k_y} \,. \notag
\end{align}
Therefore the bare Raman vertex functions for different symmetries are given by
\begin{subequations}
\begin{align}
    \gamma^{A_{1g}}_{\vk} & = t \left( \cos{k_x} + \cos{k_y} - 4 B \cos{k_x} \cos{k_y} \right) \,, \\
    \gamma^{B_{1g}}_{\vk} & = t \left( \cos{k_x} - \cos{k_y}\right) \,, \\
    \gamma^{B_{2g}}_{\vk} & = 4 t B \sin{k_x} \sin{k_y} \,.
\end{align}
\end{subequations}

The bare Raman diamagnetic bubble with the charge-charge vertices which probes the quasiparticles (QPs) response can be written as a sum over all the Brillouin zone and the Matsubara frequencies as
\begin{equation} \label{Chi33}
    \chi_{\gamma \gamma} (\vq, i \omega_n) = \dfrac{1}{\beta} \sum_{\vk} \sum_{i \nu_m} \gamma_{\vk}^i\ \gamma_{\vk}^s \ \mathrm{Tr} \big[ G^+_{\vk, m} \tau_3 G_{\vk, m} \tau_3 \big] \,,
\end{equation}
where $\gamma_{\vk}^{i,s}$ is the Raman interaction vertex for the incident/scattered light of the appropriate symmetry. Making use of the spectral representation in (\ref{spectral}) we can write the Matsubara summation in (\ref{Chi33}) as
\begin{align}
        \dfrac{1}{\beta} \sum_{i \nu_m} \mathrm{Tr} \big[ G^+_{\vk, m}  \tau_3 G_{\vk, m} \tau_3 \big] &= \dfrac{1}{\pi^2} \iint\displaylimits_{- \infty}^{+ \infty} \mathrm{d} \omega_1 \ \mathrm{d} \omega_2 \ \dfrac{1}{\beta} \sum_{i \nu_m} \dfrac{1}{i \nu_m - \omega_1} \cdot \dfrac{1}{i \nu_m + i \omega_n - \omega_2} \notag \\
        & \times \mathrm{Tr} \big[ G''(\textbf{k}, \omega_1 + i \delta) \tau_3 G''(\textbf{k}+\textbf{q}, \omega_2 + i \delta) \tau_3 \big] \,.
\end{align}
In the limit $\mathbf{q} \rightarrow 0$ we obtain the Raman susceptibility
\begin{equation}
    \begin{split}
        \chi_{\gamma \gamma}(\vq = 0, \omega) & = - \sum_{\vk}  \gamma_{\vk}^i \ \gamma_{\vk}^s \ \dfrac{\Delta_\vk^2}{E_\vk^2} \tanh{\left(\beta E_\vk/2 \right)} \left( \dfrac{1}{2 E_\vk + \omega + i \delta} + \dfrac{1}{2 E_\vk - \omega - i \delta} \right) \,,
    \end{split}
\end{equation}
which can be written using the compact notation in (\ref{average}) in the short form 
\begin{equation}
    \chi_{\gamma \gamma} (\omega) = - 2 \langle \gamma^2_{\vk} \Delta_\vk^2 \rangle \,.
\end{equation}
Following \cite{RevModPhys.79.175}, we can include the Coulomb interaction in the long-range limit giving rise to charge fluctuations (CFs) by means of the expression 
\begin{align}
       \tilde{ \chi}_{\gamma \gamma} &= \chi_{\gamma \gamma} - \dfrac{\chi_{\gamma \tau_3}^2}{\chi_{\tau_3 \tau_3}} \notag \\
        &= - 2 \langle \gamma^2_{\vk} \Delta_\vk^2 \rangle + 2 \dfrac{\langle \gamma_{\vk} \Delta_\vk^2 \rangle^2}{\langle \Delta_\vk^2 \rangle} \,,
\end{align}
with 
\begin{align}
    \chi_{\tau_3 \tau_3} (\omega) &= \dfrac{1}{\beta} \sum_{\vk'} \sum_{i \nu_m} \mathrm{Tr} \left[ G_{\vk',m}  \tau_3 G_{\vk',m}^+ \tau_3 \right] = - 2 \langle \D^2 \rangle \,, \\
    \chi_{\gamma \tau_3} (\omega) &= \dfrac{1}{\beta} \sum_{\vk'} \gamma_{\vk'} \sum_{i \nu_m} \mathrm{Tr} \left[ G_{\vk',m}  \tau_3 G_{\vk',m}^+ \tau_3 \right] = - 2 \langle \gamma_\vk \D^2 \rangle \,.
\end{align}

\subsection{C. RPA amplitude mode propagator}
The pairing interaction energy in (\ref{Hpair}) can be included by means of the RPA summation of the amplitude mode (AM) propagator
\begin{align}
    D_{\text{AM}} (\vq, \vk,\vk', \omega) = \dfrac{1}{2} \left[ V (\vk,\vk') + \sum_{\vk''} V (\vk,\vk'') \ \chi_{\tau_1 \tau_1}(\vq, \vk'', \omega) \ D_{\text{AM}} (\vq, \vk'',\vk', \omega) \right] \,,
\end{align}
where $\chi_{\tau_1 \tau_1} (\vq, \vk'', \omega)$ is the $\tau_1 - \tau_1$ bare polarization bubble. Using the usual factorization for the pairing interaction function by means of the form factor $f_\vk$, we can write $D_{\text{AM}} (\vq, \vk, \vk', \omega) = D_{\text{AM}}(\vq, \omega) f_\vk f_\vk'$ with
\begin{align}
	D_{\text{AM}} (\vq, \omega) = - \left[ 2/V + \chi_{f f} (\vq, \omega) \right]^{-1} \,. \label{Hprop}
\end{align}
\begin{figure}
    \centering
    \includegraphics[width=8cm]{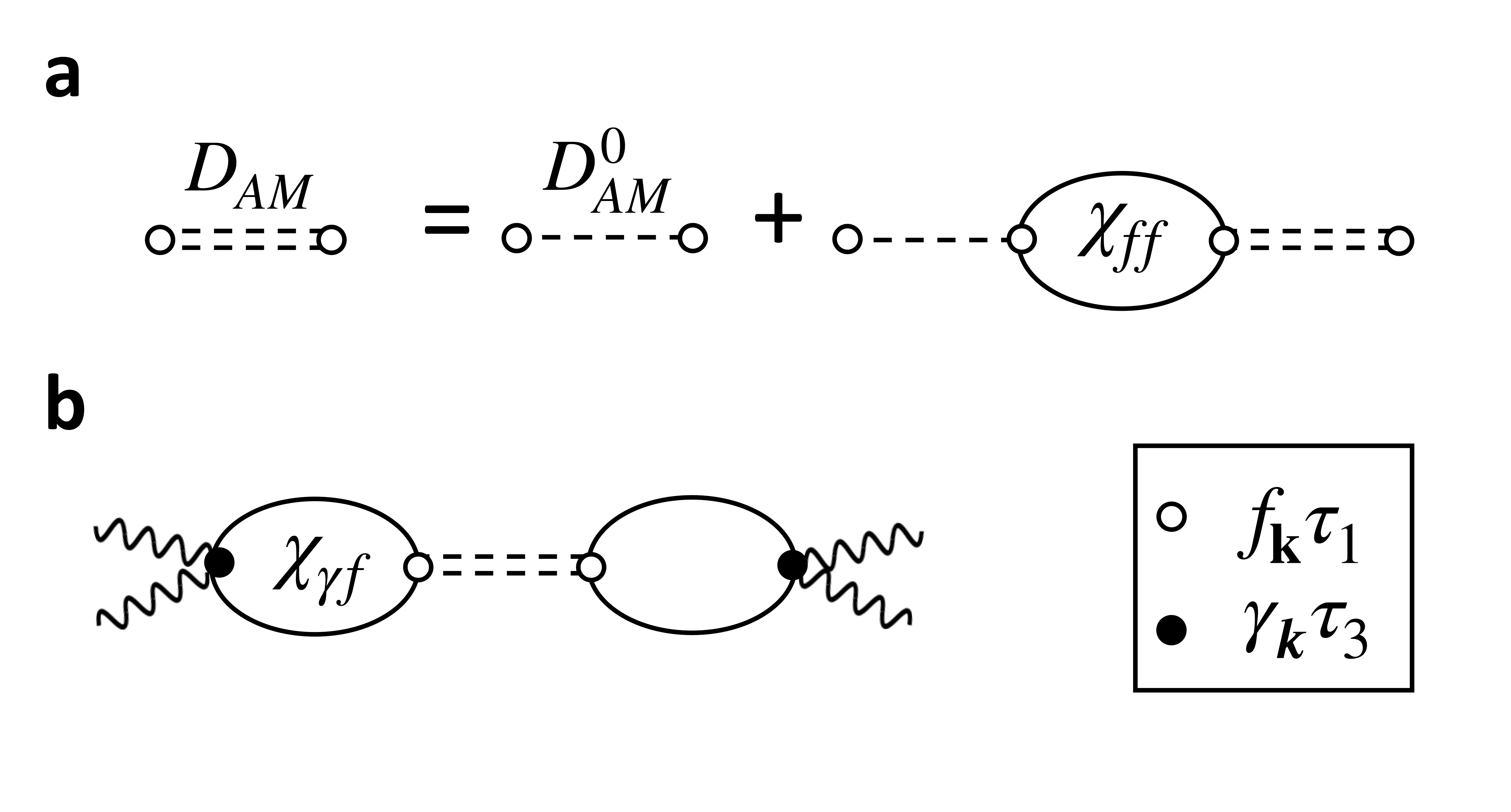}
    \caption{\textbf{Amplitude mode propagator}. (\textbf{a}) Diagrammatic representation of the dressed AM propagator $D_{\text{AM}}$ (double-dashed line) including the self-enery terms to the bare one ($D^0_{\text{AM}}$, single-dashed line) according to RPA. (\textbf{b}) Raman response with the dressed AM propagator without the many-body contribution. The inset shows the symbols used for the vertices and the corresponding vertex functions: $\circ$, amplitude channel interaction vertex $f_\vk \tau_1$; $\bullet$, Raman vertex $\gamma_\vk \tau_3$.}
    \label{Feyn-Higgs}
\end{figure}
Here we used the pairing interaction coupling in the amplitude channel with strength $V/2$ \cite{PhysRevB.93.180507}, and the $f_\vk \tau_1 - f_\vk \tau_1$ susceptibility amplitude 
\begin{align} \label{susceptibility_11}
    \chi_{f f} (\vq, \omega) &= \sum_{\vk'} f_{\vk'}^2 \ \chi_{\tau_1 \tau_1} (\vq, \vk', \omega) \notag \\
    &= \dfrac{1}{\beta} \sum_{\vk'} f_{\vk'}^2 \sum_{i \nu_m} \ \mathrm{Tr} \left[ G_{\vk',m}  \tau_1 G_{\vk',m}^+ \tau_1 \right] \,.
\end{align}
The Raman response contribution which includes the RPA AM propagator in (\ref{Hprop}) can be calculated as
\begin{align} \label{RHRPA}
    \chi_{\text{AM}} (\vq, \omega) &= \chi_{\gamma f} (\vq, \omega) \ D_{\text{AM}} (\vq, \omega) \ \chi_{f \gamma} (\vq, \omega) \notag \\
    &= - \dfrac{\chi_{\gamma f}^2 (\vq, \omega)}{2/V + \chi_{f f} (\vq, \omega)} \,,
\end{align} 
with the susceptibility $\chi_{\gamma f} (\vq, \omega) = \sum_{\vk'} f_{\vk'} \gamma_{\vk'}^i \chi_{\tau_3 \tau_1} (\vq, \vk', \omega)$, and the $\gamma_\vk \tau_3-f_\vk \tau_1$ polarization bubble 
\begin{align}
    \chi_{\tau_3 \tau_1} (\vq, \vk, \omega) &= \dfrac{1}{\beta} \sum_{i \nu_m} \mathrm{Tr} \left[ G_{\vk,m}  \tau_3 G_{\vk',m}^+ \tau_1 \right] \notag \\
    &= \iint\displaylimits_{-\infty}^{+\infty} \text{d}\omega_1 \ \text{d}\omega_2 \dfrac{ \eps \DD + \epss \D}{2 \E \EE} \ \dfrac{n_F (\omega_1) - n_F (\omega_2)}{\omega_1 - \omega_2 + \omega + i \delta} \notag \\
    &\times \left[ \delta(\omega_1 - \E) - \delta(\omega_1 + \E) \right] \left[ \delta(\omega_2 - \EE) - \delta(\omega_2 + \EE) \right] \,.
\end{align}
In the limit $\vq \rightarrow 0$ we simply get:
\begin{align}
    \chi_{\text{AM}} (\omega) &= -\dfrac{\langle 2 \gamma_\vk \eps f_\vk \D \rangle^2}{2/V - \langle 2 \eps^2 f_\vk^2 \rangle} \,.
\end{align}
Adding the CFs \cite{PhysRevB.93.180507} in Eq. (\ref{RHRPA}) we obtain 
\begin{align} \label{RHRPAsc}
    \tilde{\chi}_{\text{AM}} (\omega) &= - \dfrac{ \left( \chi_{\gamma f}(\omega) - \chi_{\gamma \tau_3} (\omega) \chi_{\tau_3 f} (\omega) / \chi_{\tau_3 \tau_3} (\omega) \right)^2}{2/V + \chi_{f f} (\omega) - \chi_{\tau_3 f}^2 (\omega) / \chi_{\tau_3 \tau_3} (\omega)} \,.
\end{align}
Defining 
\begin{subequations}
\begin{align}
    \tilde{\chi}_{\gamma f} (\omega) &= \chi_{\gamma f}(\omega) - \chi_{\gamma \tau_3} (\omega) \chi_{\tau_3 f} (\omega) / \chi_{\tau_3 \tau_3} (\omega) \,, \\
    \tilde{\chi}_{f f} (\omega) &= \chi_{f f} (\omega) - \chi_{\tau_3 f}^2 (\omega) / \chi_{\tau_3 \tau_3} (\omega) \,,
\end{align}
\end{subequations}
we can rewrite the expression in (\ref{RHRPAsc}) as
\begin{align}
    \tilde{\chi}_{\text{AM}} (\omega) &= - \dfrac{\tilde{\chi}_{\gamma f}^2 (\omega) }{2/V + \tilde{\chi}_{f f} (\omega)} \,.
\end{align}

\subsection{D. Many-body Higgs oscillations}
We now consider the many-body Higgs oscillations (HOs) by means of the vertices dressing with the pairing interaction energy terms: this allows for a mixing of the interaction channels described in section I.-A.\\
The dressed $\gamma_\vk \tau_3$ can be calculated with the Bethe-Salpeter equation in the rainbow-ladder approximation:
\begin{align} \label{GammaBS}
        \Gamma (\vk, \omega) & = \gamma_\vk \tau_3 + \dfrac{1}{\beta} \sum_{\mathbf{k'}} \sum_{i \nu_m} \dfrac{1}{2} V_{\vk, \mathbf{k'}} \ \tau_1 G^+_{\vk', m} \Gamma (\mathbf{k'}, \omega) G_{\vk',m} \tau_1 \,.
\end{align}
The simplest form that the dressed vertex function can have in order to satisfy this self-consistent equation is given by
\begin{align} \label{Gammaans}
    \Gamma (\vk, \omega) &= \Gamma_0^{(1)} (\omega) f_{\vk} \tau_1 + \left( \gamma_{\vk} + \Gamma_0^{(3)} (\omega) f_{\vk}\right) \tau_3 \,,
\end{align}
where we have neglected the spin and phase channels and putting $\Gamma_0^{(0)} = \Gamma_0^{(2)} = 0$. 
Solving the equation in Eq. (\ref{GammaBS}) using Eq. (\ref{Gammaans}) we get the coefficients
\begin{subequations}
\begin{align}
        \Gamma_0^{(1)} & = - \dfrac{\Gamma_0^{(3)} \langle f_\vk^2 \Delta_\vk \varepsilon_\vk \rangle + \langle \gamma_\vk f_\vk \Delta_\vk \varepsilon_\vk \rangle}{1/V - \langle f_\vk^2 \varepsilon_\vk^2 \rangle} \,, \\
        \Gamma_0^{(3)} & = - \dfrac{\langle \gamma_\vk f_\vk \Delta_\vk \varepsilon_\vk \rangle + 2 \langle \gamma_\vk f_\vk \Delta_\vk^2 \rangle \left( 1/V - \langle f_\vk^2 \varepsilon_\vk^2 \rangle \right)}{\langle f_\vk^2 \Delta_\vk \varepsilon_\vk \rangle + 2 \left(1/V - \langle f_\vk^2 \varepsilon_\vk^2 \rangle \right) \left(1/V + \langle f_\vk^2 \Delta_\vk^2 \rangle \right)} \,.
\end{align}
\end{subequations}
Analogously, for the dressed $\tau_1$ vertex we adopt the form:
\begin{align} \label{Lambdaans}
    \Lambda (\vk, \Omega) &= \left( 1 +\Lambda_0^{(1)} (\Omega) f_\vk \right) \tau_1 +  \Lambda_0^{(3)} (\Omega) f_\vk \tau_3 \,,
\end{align}
and solving the corresponding self-consistent equation we obtain the values
\begin{subequations}
\begin{align}
        \Lambda_0^{(1)} & = - \dfrac{\Lambda_0^{(3)} \langle f_\vk^2 \Delta_\vk \varepsilon_\vk \rangle -\langle f_\vk \varepsilon_\vk^2 \rangle}{1/V - \langle f_\vk^2 \varepsilon_\vk^2 \rangle} \,, \\
        \Lambda_0^{(3)} & = \dfrac{ \langle f_\vk \Delta_\vk \varepsilon_\vk \rangle \left(1/V - \langle f_\vk^2 \varepsilon_\vk^2 \rangle \right) + \langle  f_\vk^2 \Delta_\vk \varepsilon_\vk \rangle \langle f_\vk \varepsilon_\vk^2 \rangle}{\langle f_\vk^2 \Delta_\vk \varepsilon_\vk \rangle^2 + \left(1/V - \langle f_\vk^2 \varepsilon_\vk^2 \rangle \right) \left(1/V + \langle f_\vk^2 \Delta_\vk^2 \rangle \right)} \,.
\end{align}
\end{subequations}
\begin{figure}
    \centering
    \includegraphics[width=8cm]{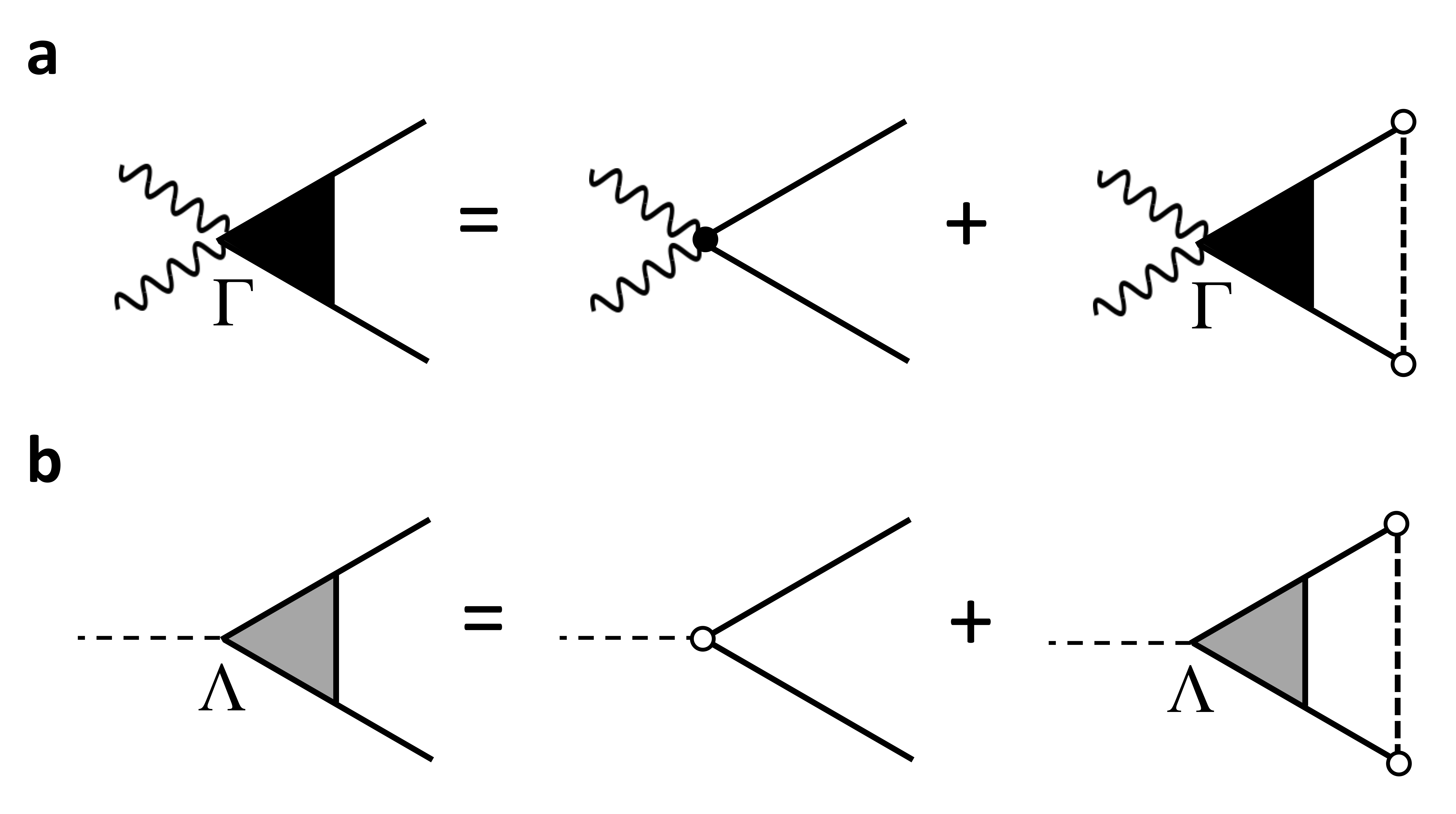}
    \caption{\textbf{Many-body Higgs oscillations}. Diagrammatic representation of the sum of many-body Higgs oscillations (within the ladder approximation) for the Raman vertex $\gamma_\vk \tau_3$ (\textbf{a}) and the amplitude interaction vertex $f_\vk \tau_1$ (\textbf{b}), as expressed in Eq. (\ref{Gammaans}) and Eq. (\ref{Lambdaans}), respectively. With the replacements $\gamma_\vk \tau_3 \rightarrow \Gamma$ and  $\tau_1 \rightarrow \Lambda$ in Fig. ~\ref{Feyn-Higgs}(\textbf{b}) the full many-body Higgs oscillations contribution to the Raman response is obtained.}
    \label{VC}
\end{figure}
Thus, the full Raman response obtained using the dressed vertices is written as
\begin{align}
    \tilde{\chi}_{\text{full}} = \tilde{\chi}_{\Gamma \Gamma} - \dfrac{(\tilde{\chi}_{\Gamma \Lambda})^2 }{2/V + \tilde{\chi}_{\Lambda \Lambda}} \,,
\end{align}
where the susceptibilities screened by the long-range Coulomb interaction are given by
\begin{subequations}
\begin{align}
    \tilde{\chi}_{\Gamma \Gamma} &= \chi_{\Gamma \Gamma} - \chi_{\Gamma \tau_3}^2 / \chi_{\tau_3 \tau_3} \,, \\
    \tilde{\chi}_{\Gamma \Lambda} &= \chi_{\Gamma \Lambda} - \chi_{\Gamma \tau_3} \ \chi_{\tau_3 \Lambda} / \chi_{\tau_3 \tau_3} \,, \\
    \tilde{\chi}_{\Lambda \Lambda} &= \chi_{\Lambda \Lambda} - \chi_{\Lambda \tau_3}^2 / \chi_{\tau_3 \tau_3} \,.
\end{align}
\end{subequations}
We used the definitions of the unscreened susceptibilities as follows:
\begin{subequations}
\begin{align}
    \chi_{\Gamma \Gamma} &= \dfrac{1}{\beta} \sum_{\vk'} \sum_{i \nu_m} \mathrm{Tr} \left[ G_{\vk',m} \ \Gamma (\vk', \Omega) \ G^+_{\vk',m} \ \Gamma (\vk', \Omega)  \right] \,, \\
    \chi_{\Lambda \Lambda} &= \dfrac{1}{\beta} \sum_{\vk'} \sum_{i \nu_m} \mathrm{Tr} \left[ G_{\vk',m} \ f_{\vk'} \Lambda (\vk', \Omega) \ G^+_{\vk',m} \ f_{\vk'} \Lambda (\vk', \Omega)  \right] \,, \\
    \chi_{\Gamma \Lambda} &= \dfrac{1}{\beta} \sum_{\vk'} \sum_{i \nu_m} \mathrm{Tr} \left[ G_{\vk',m} \ \Gamma (\vk', \Omega) \ G^+_{\vk',m} \ f_{\vk'} \Lambda (\vk', \Omega) \right] \,. 
\end{align}
\end{subequations}
Their full expressions in term of averages are given by:
\begin{subequations}
\begin{align}
        \chi_{\Gamma \Gamma} (\Omega) & =  -2 \langle \gamma_\vk^2 \D^2 \rangle + 2 \Gamma_0^{(1)} \langle \gamma_\vk f_\vk \D \eps \rangle - 2 \Gamma_0^{(3)} \langle \gamma_\vk f_\vk \D^2 \rangle + 2 \Gamma_0^{(1)} \Gamma_0^{(3)} \langle f_\vk^2 \D \eps \rangle \notag \\
        & - \left( \Gamma_0^{(1)} \right)^2 \langle f^2_\vk \eps^2 \rangle - \left( \Gamma_0^{(3)} \right)^2 \langle f^2_\vk \D^2 \rangle \,, \\
        \chi_{\Lambda \Lambda} (\Omega) & = -2 \langle f^2_\vk \varepsilon_\vk^2 \rangle - 2 \Lambda_0^{(1)} \langle f^3_\vk \varepsilon_\vk^2 \rangle + 2 \Lambda_0^{(3)} \langle f^3_\vk \Delta_\vk \varepsilon_\vk \rangle + 2 \Lambda_0^{(1)} \Lambda_0^{(3)} \langle f^4_\vk \Delta_\vk \varepsilon_\vk \rangle \notag \\
        & - \left( \Lambda_0^{(1)} \right)^2 \langle f^4_\vk \varepsilon_\vk^2 \rangle - \left( \Lambda_0^{(3)} \right)^2 \langle f^4_\vk \Delta_\vk^2 \rangle \\
        \chi_{\Gamma \Lambda} (\Omega) & = 2 \langle \gamma_\vk f_\vk \Delta_\vk \varepsilon_\vk \rangle - \Gamma_0^{(1)} \langle f^2_\vk \varepsilon_\vk^2 \rangle + \Gamma_0^{(3)} \langle f^2_\vk \Delta_\vk \varepsilon_\vk \rangle + \Lambda_0^{(1)} \langle f^2_\vk \gamma_\vk \Delta_\vk \varepsilon_\vk \rangle \,, \notag \\
        &- \Lambda_0^{(1)} \Gamma_0^{(1)} \langle f^3_\vk \varepsilon_\vk^2 \rangle + \Lambda_0^{(1)} \Gamma_0^{(3)} \langle f^3_\vk \Delta_\vk \varepsilon_\vk \rangle - \Lambda_0^{(3)} \langle \gamma_\vk f^2_\vk \Delta_\vk^2 \rangle + \Lambda_0^{(3)} \Gamma_0^{(1)} \langle f^3 \Delta_\vk \varepsilon_\vk \rangle - \Lambda_0^{(3)} \Gamma_0^{(3)} \langle f^3_\vk \Delta_\vk^2 \rangle \,.
\end{align}
\end{subequations}

\section{II. Phonon-Higgs contribution}
\subsection{A. Phonon-Higgs Raman response}
Let's consider a phonon with $\mathbf{q} = 0$ and frequency $\omega_0$: its bare propagator is given by
\begin{align}
    D_{\text{ph}}^0 (\Omega) &= \dfrac{- 2 \omega_0}{\omega_0^2 - (\Omega + i \delta)^2} \,.
\end{align}
Including the screening due to the RPA electron-phonon interaction, with $g_D (\vk) = g_0 g_\vk$ being the electron-phonon coupling function specific of a given phonon symmetry, we get the dressed propagator $D_{\text{ph}} (\Omega) = D_{ph}^0 (\Omega) + g_0^2 D_{\text{ph}}^0 (\Omega) \chi_{gg} (\Omega) D_{\text{ph}}(\Omega)$, thus
\begin{align}
    D_{\text{ph}} (\Omega) &= \dfrac{- 2 \omega_0}{\omega_0^2 + 2 g_0^2 \omega_0 \chi_{gg} (\Omega) - (\Omega + i \delta)^2} \,,
\end{align}
with 
\begin{align}
    \chi_{gg} (\Omega) &= \sum_\vk g_\vk^2 \chi_{33} (\vk, \Omega) \,.
\end{align}
This can be calculated for the superconducting state, obtaining
\begin{align}
    \chi_{gg} (\Omega) &=  \sum_\vk g_\vk^2 \dfrac{- 4 \Delta_\vk^2}{\E \left[ 4 \E^2 - (\Omega + i \delta)^2 \right]} \,.
\end{align}
Below $T_c$, we can also include interaction with the AM propagator, obtained via RPA summation, $\tilde{D} (\Omega) = D_{\text{ph}} (\Omega) + g_0^2 D_{\text{ph}}(\Omega) D_{\text{AM}}(\Omega) \chi_{gf}^2 (\Omega) \tilde{D}_{\text{ph}} (\Omega)$, getting 
\begin{align}
    \tilde{D}_{\text{ph}} (\Omega) &= \dfrac{D_{\text{ph}} (\Omega)}{1 + g_0^2 D_{\text{ph}}(\Omega) \dfrac{\chi_{gf}^2 (\Omega)}{2/V + \chi_{ff} (\Omega)}} \notag \\
    &= \dfrac{-2 \omega_0}{\omega_0^2 + 2 g_0^2 \omega_0 \chi_{gg} (\Omega) - 2 g_0^2 \omega_0 \dfrac{\chi_{gf}^2 (\Omega)}{2/V + \chi_{ff} (\Omega)} - (\Omega + i \delta)^2} \,,
\end{align}
with 
\begin{align}
    \chi_{gf} (\Omega) &= \sum_\vk g_\vk f_\vk \dfrac{4 \eps \Delta_\vk}{\E \left[ 4 \E^2 - (\Omega + i \delta)^2 \right]} \,, \\
    \chi_{ff} (\Omega) &= \sum_\vk f_\vk^2 \dfrac{- 4 \eps^2}{\E \left[ 4 \E^2 - (\Omega + i \delta)^2 \right]} \,.
\end{align}
The phononic Raman susceptibility is given by 
\begin{align}
    \chi_{\text{ph}} (\Omega) &= g^2 \chi_{\gamma g}^2 (\Omega) \tilde{D}_{ph}(\Omega) \,,
\end{align}
with 
\begin{align}
    \chi_{\gamma g} &= \sum_\vk g_\vk \gamma_\vk \dfrac{- 4 \Delta_\vk^2}{\E \left[ 4 \E^2 - (\Omega + i \delta)^2 \right]} \,, \\
\end{align}
with $\gamma_\vk$ the Raman vertex.
\begin{figure}[t!]
    \centering
    \includegraphics[scale=0.7]{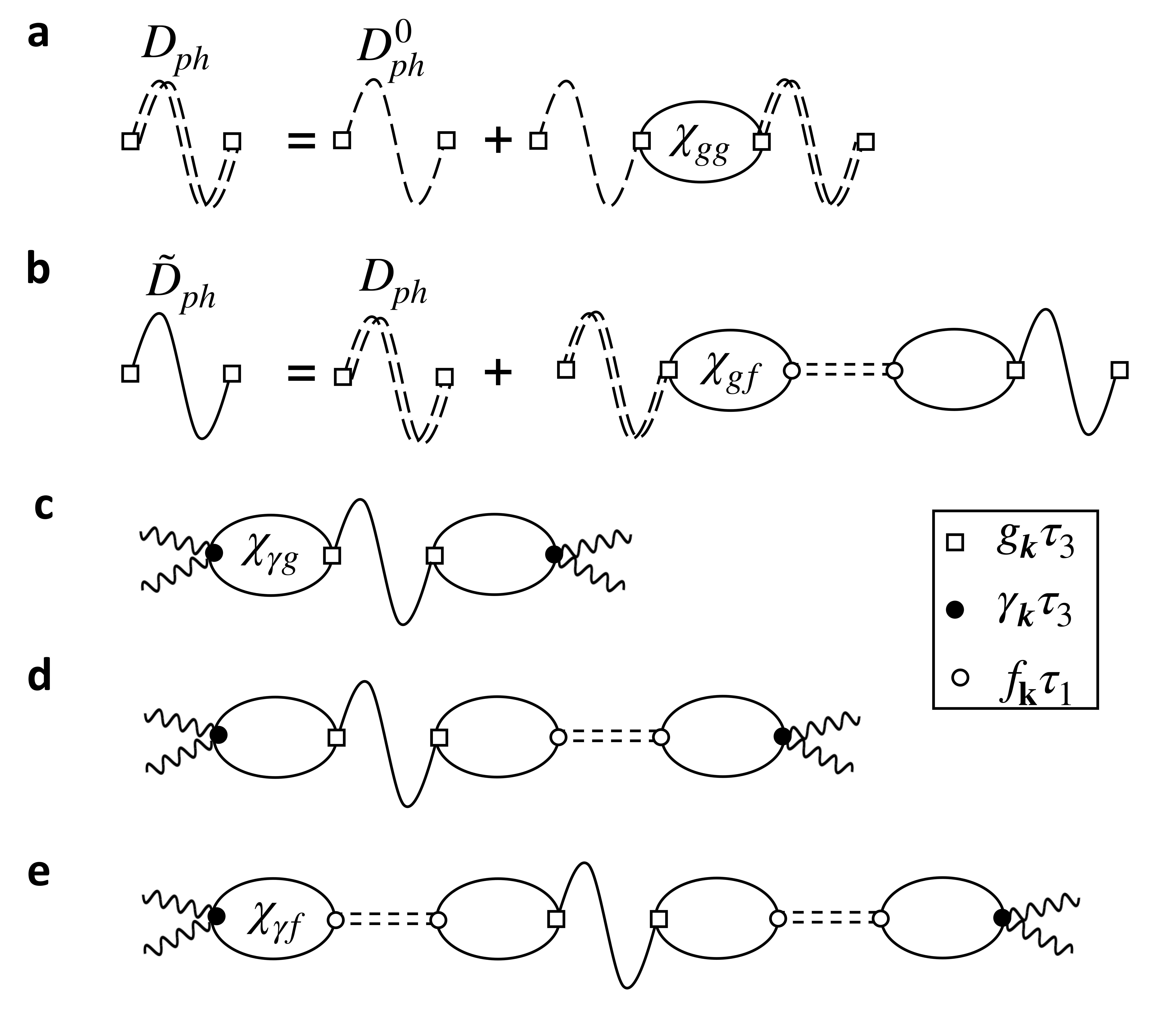}
    \caption{\textbf{Diagrammatic summation of phonon and amplitude mode contributions}. Diagrammatic representation of the Raman response of a phonon (\textbf{c}). The phonon propagator is dressed with the quasiparticles self-energy (\textbf{a}) and the amplitude mode (\textbf{b}) via a self-consistent RPA summation. (\textbf{e}), (\textbf{f}) represent the mixed dressed phonon-AM interaction terms.}
    \label{phonons1}
\end{figure}\\
We can also add the explicit interaction of the phonon with the AM in the Raman response (Fig. ~\ref{phonons1} \textbf{c}-\textbf{e}), obtaining the total phononic contribution
\begin{align} \label{phononTOT}
    \chi_{\text{ph, ~tot}}(\Omega) &= \tilde{D}_{\text{ph}}(\Omega) g^2 \left[ \chi_{\gamma g}(\Omega) + \chi_{\gamma f}(\Omega) \chi_{g f}(\Omega) D_{AM} (\Omega) \right]^2 \notag \\
    &= \dfrac{-2 \omega_0  g^2 \bigg[ \chi_{\gamma g} - \dfrac{\chi_{\gamma f} \chi_{g f}}{{2/V + \chi_{ff} }} \bigg]^2}{\omega_0^2 + 2 g_0^2 \omega_0 \chi_{gg} - (\Omega + i \delta)^2 - 2 g_0^2 \omega_0 \chi_{gf}^2/(2/V + \chi_{ff})}
\end{align}
We can now include the CFs due to Coulomb interaction in the long wavelength limit ($q \rightarrow 0$ in the Coulomb potential $4 \pi e^2/q$) using the following functions rather than the bare ones:
\begin{align}
    \tilde{\chi}_{\gamma g} &= \chi_{\gamma g} - \dfrac{\chi_{\gamma \tau_3} \chi_{g \tau_3}}{\chi_{\tau_3 \tau_3}} = -2 \left[ \langle \gamma_\vk g_\vk \D^2 \rangle - \dfrac{\langle \gamma_\vk \D^2 \rangle \langle g_\vk \D^2 \rangle}{\langle \D^2 \rangle} \right] \,, \\
    \tilde{\chi}_{\gamma f} &= \chi_{\gamma f} - \dfrac{\chi_{\gamma \tau_3} \chi_{f \tau_3}}{\chi_{\tau_3 \tau_3}} = 2 \left[ \langle \gamma_\vk f_\vk \eps \D \rangle - \dfrac{\langle \gamma_\vk \D^2 \rangle \langle \eps f_\vk \D \rangle}{\langle \D^2 \rangle} \right]  \,, \\
    \tilde{\chi}_{g f} &= \chi_{g f} - \dfrac{\chi_{g \tau_3} \chi_{f \tau_3}}{\chi_{\tau_3 \tau_3}} = 2 \left[ \langle g_\vk f_\vk \eps \D \rangle - \dfrac{\langle g_\vk \D^2 \rangle \langle f_\vk \eps \D \rangle}{\langle \D^2 \rangle} \right] \,, \\
    \tilde{\chi}_{gg} &= \chi_{g g} - \dfrac{\chi_{g \tau_3}^2}{\chi_{\tau_3 \tau_3}} = -2 \left[ \langle g_\vk^2 \D^2 \rangle - \dfrac{\langle g_\vk \D^2 \rangle^2}{\langle \D^2 \rangle} \right] \,, \\
    \tilde{\chi}_{ff} &= \chi_{f f} - \dfrac{\chi_{f \tau_3}^2}{\chi_{\tau_3 \tau_3}} = - 2 \left[ \langle \eps^2 f_\vk^2 \rangle - \dfrac{\langle f_\vk \eps \D \rangle^2}{\langle \D^2 \rangle} \right] \,.
\end{align}

\subsection{B. Phonons with many-body Higgs interactions}
We now include the interaction of the phonons with many-body Higgs oscillations by means of the vertex corrections in the very same way as used for the Raman response calculated in Section I.D.\\
Thus, the electron-phonon interaction vertex $g_\vk$ for a given symmetry gets dressed $g_\vk \tau_3 \rightarrow \Phi_\vk$, defined as $\Phi_\vk = \Phi_0^{(1)} f_\vk \tau_1 + (g_\vk + \Phi_0^{(3)} f_\vk) \tau_3$ calculated as follows:
\begin{align}
    \Phi_0^{(1)} &= \dfrac{-\Phi_0^{(3)} \langle f_\vk^2 \D \eps \rangle + \langle g_\vk f_\vk \D \eps \rangle}{1/V - \langle f_\vk^2 \eps^2 \rangle} \,, \\
    \Phi_0^{(3)} &= - \dfrac{\langle g_\vk f_\vk \D \eps \rangle + 2 \langle g_\vk f_\vk \D^2 \rangle \left( 1/V - \langle f_\vk^2 \eps^2 \rangle \right)}{\langle f_\vk^2 \D \eps \rangle + 2 \left( 1/V - \langle f_\vk^2 \eps^2 \rangle \right) \left( 1/V + \langle f_\vk^2 \D^2 \rangle \right)} \,.
\end{align}
Then we rewrote Eq. (\ref{phononTOT}) using the dressed vertices with many-body Higgs interactions, namely obtaining
\begin{align} \label{phononVC}
    \chi_{\text{ph, HO}}(\Omega) &= \tilde{D}_{\text{ph}}(\Omega) \ g^2 \left[ \tilde{\chi}_{\Gamma \Phi}(\Omega) + \tilde{\chi}_{\Gamma \Lambda}(\Omega) \tilde{\chi}_{\Phi \Lambda}(\Omega) \tilde{D}_{\text{AM}} (\Omega) \right]^2 \,,
\end{align}
where we also took into account of the CFs due to Coulomb long-range interactions into the expressions of the susceptibilities and the phonon and AM propagators.

\section{III. Integration around the Fermi Surface} 
The tight-binding band dispersion used so far is derived from a reduction of a three band model which gives a large contribution to the density of states at the Fermi level for cuprates. However, in order to take into account of admixtures of more Fermi surface harmonics to the Raman contributions for different symmetries, we adopted a calculation independent of the details of the band dispersion, namely integrating the response functions around the Fermi surface and including higher harmonics.

\subsection{A. Integrals around the Fermi surface}
Let's assume that the energy band dispersion $\varepsilon_\vk$ depends only on the absolute value of $\vk$, while the form factor given by the gap symmetry depends only on the polar angle $f_\vk = f(\varphi)$ (and thus $\Delta_\vk \rightarrow \Delta (\varphi) = \Delta_{\text{max}} f(\varphi)$). We can replace the summations over the points $\vk$ to integrals over the energy $\varepsilon$ (around the Fermi level in the interval $[- \varepsilon_c, + \varepsilon_c]$, $\varepsilon_c$ being the appropriate cut-off energy) and the azimuthal angle $\varphi$:
\begin{align}
    V \sum_\vk \rightarrow V \int\displaylimits_{0}^{2 \pi} \text{d}\varphi \int\displaylimits_{- \varepsilon_c}^{+ \varepsilon_c} \text{d}\varepsilon \ D(\varepsilon) \,,
\end{align}S
where $D(\varepsilon)$ is the density of states, such that $\int\displaylimits_{- \varepsilon_c}^{+ \varepsilon_c} D(\varepsilon) \ \text{d}\varepsilon = 1$. We assume that the density of states is constant around the Fermi energy, $D(0) \approx const. = 1/(2 \varepsilon_c)$. Thus we can write 
\begin{align}
    V \sum_\vk \rightarrow V D(0) \int\displaylimits_{0}^{2 \pi} \text{d}\varphi \int\displaylimits_{- \varepsilon_c}^{+ \varepsilon_c} \text{d}\varepsilon \,.
\end{align}
We can define $\lambda \equiv V D(0)$ to shorten the notation. In general we can substitute to the discrete sum over the $\vk$ points the integral $\sum_\vk \rightarrow D(0) \int\displaylimits_{0}^{2 \pi} \text{d}\varphi \int\displaylimits_{- \varepsilon_c}^{+ \varepsilon_c} \text{d}\varepsilon$. Making use of the gap equation we can calculate the value of $\lambda$ ($V$):
\begin{align}
    \lambda \int\displaylimits_{0}^{2 \pi} \text{d}\varphi \int\displaylimits_{- \varepsilon_c}^{+ \varepsilon_c} \text{d}\varepsilon \dfrac{f(\varphi)^2}{2 \sqrt{\varepsilon^2 + \Delta_{\max}^2 f(\varphi)^2}}  = 1 \,.
\end{align}
So that we obtain
\begin{align}
    \lambda = \left[ \int\displaylimits_{0}^{2 \pi} \text{d}\varphi \ f(\varphi)^2 \ln{\left(\dfrac{2 \varepsilon_c}{\Delta_{max} |f(\varphi)|}\right)} \right]^{-1} \ \,. \label{lambda_c}
\end{align}
We observe that this value depends upon the energy cut-off $\varepsilon_c$, so that $\lambda = \lambda(\varepsilon_c)$. In the limit $\varepsilon_c \gg \Delta_{max}$ we can substitute the integral $\int\displaylimits_{- \varepsilon_c}^{+ \varepsilon_c} \text{d} \varepsilon$ with $\int\displaylimits_{- \infty}^{+ \infty} \text{d} \varepsilon$. Instead, if we consider only the Fermi surface, i.e. the density of states being $D(\varepsilon) = \delta(0)$, then we would get
\begin{align}
    \lambda = \left[ \int\displaylimits_{0}^{2 \pi} \text{d}\varphi \dfrac{|f(\varphi)|}{2\Delta_{\max}} \right]^{-1} \ \,,
\end{align}
which does not depend upon any cut-off.\\
We introduce now for the Fermi-surface integration approach the definition of the average $\langle x \rangle$ analogous to the one in (\ref{average}) as follows:
\begin{align}
    \langle x(\varepsilon, \varphi) \rangle \equiv \int\displaylimits_0^{2 \pi} \text{d} \varphi \int\displaylimits_{- \infty}^{+ \infty} \text{d}\varepsilon \ \dfrac{4 x(\varepsilon, \varphi)}{2 E (4 E^2 - \Omega^2)} \,.
\end{align}
If $x$ depends only on the polar angle, then we can solve the integral over the energy and get
\begin{align}
    \langle x(\varphi) \rangle = \int\displaylimits_0^{2 \pi} \text{d} \varphi \ \dfrac{x(\varphi)}{\Omega \sqrt{4 \Delta^2(\varphi) - \Omega^2}} \arcsin{\left( \dfrac{\Omega}{2|\Delta (\varphi)|} \right)} \,.
\end{align}

\subsection{B. Raman response}
We consider a $d$-wave superconductor with $D_{4h}$ lattice point group symmetry, and we express the gap function depending only on the azimuthal angle $\varphi$ as $\Delta (\varphi) = \Delta_{\text{max}} f(\varphi)$, with the form factor $f(\varphi) = \cos{(2 \varphi)}$. We first consider a cylindrical Fermi surface, and then we allow for deviations from cylindricity retaining the lowest terms in a Fermi-surface harmonic expansion depending on azimuthal angle $\varphi$ \cite{PhysRevB.51.16336}:
\begin{subequations}
\begin{align}
    \gamma_{A_{1g}} &= \alpha_0 \ \varepsilon + \alpha_1 \cos{(4 \varphi)} \,, \\
    \gamma_{B_{1g}} &= \beta_1 \cos{(2 \varphi)} = \beta_1 f(\varphi) \,, \\
    \gamma_{B_{2g}} &= \gamma_1 \sin{(2 \varphi)} \,,
\end{align}
\end{subequations}
Where $\alpha_0$, $\alpha_1$, $\alpha_2$ and $\beta_1 = \gamma_1$ are parameters related to the material. If $\alpha_0 = 0$ we recover the result of the integral over the Fermi circle: in this case no AM is present, because all the integrals over $\varepsilon$ are odd and go to zero. If we set $\alpha_1 = 0$ and $\alpha_0 = -1/2$ (which is the case for a square lattice, or $B = 0$ referring to the tight-binding model in (\ref{dispersion})), then we are in the circular Fermi surface limit. In the general case with $\alpha_0 = -1/2$ and $\alpha_1 \neq 0$ we obtain the full result \cite{PhysRevB.92.064508, PhysRevB.101.184519}. We can also add higher harmonics to the $A_{1g}$ vertex \cite{PhysRevB.51.16336}, obtaining the final expression:
\begin{align}
    &\gamma_{A_{1g}} = \alpha_0 \ \varepsilon + \alpha_1 \cos{(4 \varphi)} + \alpha_2 \cos{(8 \varphi)} \,.
\end{align}
The bare $A_{1g}$ Raman response without AM and with CFs due to long-range Coulomb interaction is given by
\begin{align}
    \tilde{\chi}_{A_{1g}}^{\text{QP+CF}} & = \chi_{\gamma \gamma} - \chi_{\gamma \tau_3}^2/\chi_{\tau_3 \tau_3} \notag \\
    & = -2 \langle \gamma_{A_{1g}}^2 (\varepsilon, \varphi) \Delta^2(\varphi) \rangle + 2 \dfrac{\left( \langle \gamma_{A_{1g}} (\varepsilon, \varphi) \Delta^2(\varphi) \rangle \right)^2}{\langle \Delta^2(\varphi) \rangle} \,.
\end{align}
Substituting the expressions for the vertex and the gap function, retaining only the nonzero integrals we get
\begin{align}
    \tilde{\chi}_{A_{1g}}^{\text{QP+CF}} & = -2 \Delta_{max}^2 \bigg[ \alpha_0^2 \langle f^2(\varphi) \varepsilon^2 \rangle + \alpha_1^2 \langle f^2(\varphi) \cos{(4 \varphi)}^2 \rangle + \alpha_2^2 \langle f^2(\varphi) \cos{(8 \varphi)}^2  \rangle \notag \\
    & + \alpha_1 \alpha_2 \langle f^2(\varphi) \cos{(8 \varphi)} \cos{(4 \varphi)}  \rangle \bigg] + 2 \Delta_{max}^2 \bigg[ \alpha_1^2 \langle f^2(\varphi) \cos{(4 \varphi)} \rangle^2 + \alpha_2^2 \langle f^2(\varphi) \cos{(8 \varphi)} \rangle^2 \notag \\ 
    & + \alpha_1 \alpha_2 \langle f^2(\varphi) \cos{(4 \varphi)} \rangle \langle f^2(\varphi) \cos{(8 \varphi)} \rangle \bigg]/\langle \cos{(2 \varphi)}^2 \rangle \,.
\end{align}
The amplitude mode contribution to the $A_{1g}$ response calculated within the RPA is given by
\begin{align}
    \chi^{\text{AM}}_{A_{1g}} & = - \dfrac{\langle 2 \gamma_{A_{1g}} \varepsilon f(\varphi) \Delta(\varphi) \rangle^2}{2/\lambda - 2 \langle f^2(\varphi) \varepsilon^2 \rangle} = - 2 \Delta_{\text{max}}^2 \dfrac{\langle f^2(\varphi) \varepsilon^2 \rangle^2}{1/\lambda - \langle f^2(\varphi) \varepsilon^2 \rangle} \,.
\end{align}
The $B_{1g}$ and $B_{2g}$ responses are calculated as
\begin{align}
    \chi_{B_{1g}} & = - 2 \langle \gamma_{B_{1g}}^2 \Delta^2(\varphi) \rangle = - 2 \beta_1^2 \Delta^2_{\text{max}} \langle f^4(\varphi) \rangle \,, \\
    \chi_{B_{2g}} & = - 2 \langle \gamma_{B_{2g}}^2 \Delta^2(\varphi) \rangle = - 2 \gamma_1^2 \Delta^2_{\text{max}} \langle \sin^2(2 \varphi) f^2(\varphi) \rangle \,.
\end{align}

\subsection{C. Many-body Higgs contribution}
Considering the many-body effect of Higgs oscillations through the vertex corrections, for the $A_{1g}$ symmetry we get the susceptibilities:
\begin{subequations}
\begin{align}
    \chi_{\Gamma \Gamma} &= - 2 \Delta_{\text{max}}^2 \left[ \alpha_0^2 \langle f^2 \varepsilon^2 \rangle + \alpha_1^2 \langle f^2 \cos^2{(4 \varphi) \rangle + \alpha_2^2 \langle f^2 \cos^2{(8 \varphi)} + 2 \alpha_1 \alpha_2 \langle f^2 \cos{(4 \varphi) \cos{(8 \varphi)}} \rangle \rangle} \right] \notag \\
    &+ 2 \Gamma_0^{(1)} \Delta_{\text{max}} \alpha_0 \langle f^2 \varepsilon^2 \rangle - (\Gamma_0^{(1)})^2  \langle f^2 \varepsilon^2 \rangle - 2 \Gamma_0^{(3)} \Delta_{\text{max}}^2 \left( \alpha_1 \langle f^3 \cos{(4 \varphi)} \rangle + \alpha_2 \langle f^3 \cos{(8 \varphi)} \rangle \right) \notag \\
    &- (\Gamma_0^{(3)})^2 \Delta_{\text{max}}^2 \langle f^4 \rangle \,, \\
    \chi_{\Gamma \Lambda} &= \left( 2 \Delta_{\text{max}} \alpha_0 - \Gamma_0^{(1)} + \Lambda_0^{(1)} \Delta_{\text{max}} \alpha_0 - \Lambda_0^{(1)} \Gamma_0^{(1)} \right) \langle f^2 \varepsilon^2 \rangle \,, \\
    \chi_{\Lambda \Lambda} &= -2 \langle f^2 \varepsilon^2 \rangle - 2 \Lambda_0^{(1)} \langle f^3 \varepsilon^2 \rangle - (\Lambda_0^{(1)})^2 \langle f^4 \varepsilon^2 \rangle \,.
\end{align}
\end{subequations}
The vertices are given by
\begin{subequations}
\begin{align}
    \Gamma_0^{(1)} &= - \dfrac{\Delta_{\text{max}} \alpha_0 \langle f^2 \varepsilon^2 \rangle}{1/\lambda - \langle f^2 \varepsilon^2 \rangle} \,, \\
    \Gamma_0^{(3)} &= - \dfrac{\Delta_{\text{max}} \alpha_0 \langle f^2 \varepsilon^2 \rangle + 2 \Delta_{\text{max}}^2  \left(\langle \alpha_1 \langle f^3\cos{(4 \varphi)} \rangle + \alpha_2 \langle f^3 \cos{(8 \varphi)} \rangle \right) \left( 1/\lambda - \langle f^2 \varepsilon^2 \rangle \right) }{2 \left( 1/\lambda - \langle f^2 \varepsilon^2 \rangle \right) \left( 1/\lambda + \Delta_{\text{max}}^2 \langle f^4 \rangle \right)} \,, \\
    \Lambda_0^{(1)} &= \dfrac{\langle f \varepsilon^2 \rangle}{1/\lambda - \langle f^2 \varepsilon^2 \rangle} \,,
\end{align}
\end{subequations}
while $\Lambda_0^{(3)} = 0$. \\
On the other hand, for the $B_{1g}$ response we have:
\begin{align}
    \chi_{\Gamma \Gamma} &= - \Delta_{\text{max}}^2 \left(2 \beta_1^2  + 2 \beta_1 \Gamma_0^{(3)} +  (\Gamma_0^{(3)})^2 \right) \langle f^4 \rangle \,, \label{chi_gg_B1g}
\end{align}
while $\chi_{\Gamma \Lambda} = 0 $, with 
\begin{subequations}
\begin{align}
    \Gamma_0^{(3)} &= - \dfrac{\beta_1 \Delta_{\text{max}}^2 \langle f^4 \rangle}{1/\lambda + \Delta_{\text{max}}^2 \langle f^4 \rangle} \,, \\
    \Lambda_0^{(1)} &= \dfrac{\langle f \varepsilon^2 \rangle}{1/\lambda - \langle f^2 \varepsilon^2 \rangle} \,,
\end{align}
\end{subequations}
and $\Gamma_0^{(1)} = \Lambda_0^{(3)} = 0$. Here we used the value of $\lambda$ as given in Eq. (\ref{lambda_c}), with the pairing interaction strength $\lambda/2$. For the $B_{2g}$ Raman susceptibility it is enough to replace $\beta_1 f(\varphi) \rightarrow \gamma_1 \sin(2 \varphi)$ in Eq. (\ref{chi_gg_B1g}).\\
Then, we can add the contributions of CFs, analogously to the previous case.

\section{IV. Impurities and $eV$-light Raman}
As shown by N. Tsuji and Y. Nomura \cite{PhysRevResearch.2.043029}, the resonant paramagnetic superconducting response which is activated by impurities is given by the following diagrams:

\begin{figure}[h!]
\includegraphics[scale=0.5]{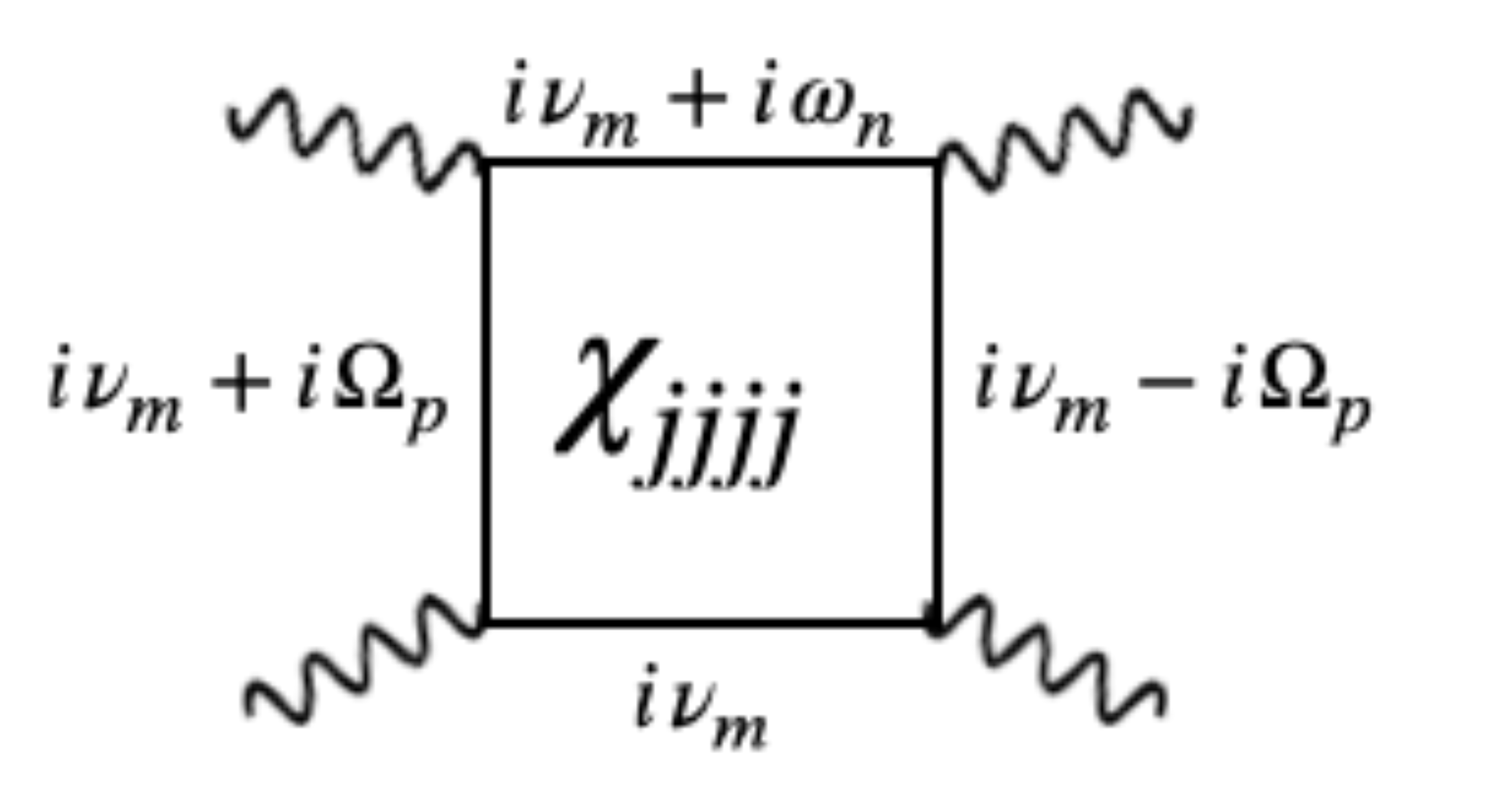}
\end{figure}
\begin{figure}[h!]
\includegraphics[scale=0.5]{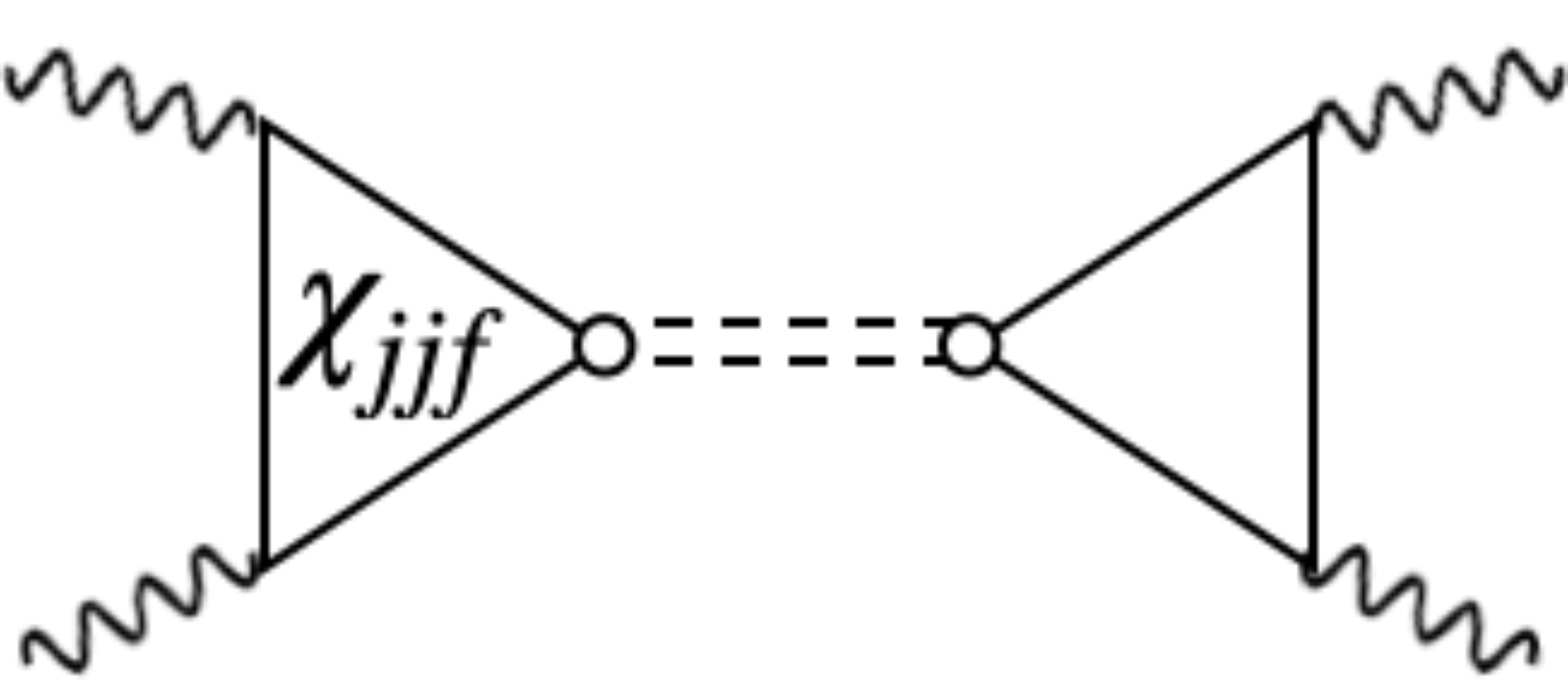}
\end{figure}

Without writing explicitly the $\vk$-dependence of the propagators and  the momentum dependence due to the impurity scattering, the quasiparticles' paramagnetic susceptibility is given by  the expression
\begin{align}
\chi_{jjjj} (\omega_R,  \Omega_{eV}) =\lim_{i \omega_n \rightarrow \omega_R} \ \lim_{i \Omega_p \rightarrow \Omega_{eV} } \  \dfrac{1}{\beta} \sum_{i \nu_m} j^4 \ \mathrm{Tr} \lbrace G (i \nu_m) G (i \nu_m + i \Omega_p) G (i \nu_m + i \omega_n) G (i \nu_m - i \Omega_p) \rbrace \,, 
\end{align}
where $\Omega_{eV}$ and $\omega_R$ are the eV-light frequency and the Raman frequency shift, respectively.  This susceptibility vanishes exactly in the superconducting clean limit \cite{PhysRevB.93.180507}, while it is activated (i.e.: becomes  non-zero) in the presence of impurities. The Green's function expressed in a matricial form in Nambu-Gor'kov space reads:
\begin{align}
\lim_{i \Omega_p \rightarrow \Omega_{eV}} \ G (i \nu_m \pm i \Omega_p) = \left[ i \nu_m \pm \Omega_{eV} - \eps \tau_3 - \D \tau_1 \right]^{-1} \,, 
\end{align}
where $\tau_i$ are the Pauli matrices. From this expression we can realize that the paramagnetic process is in an off-resonant condition, since the pole of the fermionic propagator $G$ is shifted up to the eV range by the light energy $\Omega_{eV}$, far away from the resonance of the superconducting gap. \\
The same arguments are valid for the diagram inclluding the Higgs mode, which involves the susceptibility $\chi_{jjf}$.\\
Therefore, in the non-resonant eV-light Raman processes the impurity contribution \citep{}, which comes mainly from the processes described which are off-resonant in this regime, can be fairly neglected.

\section{V. Spin susceptibility, spin-charge and spin-Higgs coupling}
The spin susceptibility for cuprates is given by the following expression \cite{PhysRevLett.94.147001,PhysRevB.75.184534}:
\begin{align}
\chi (\vq, \omega) = \dfrac{\chi_0 (\vq, \omega)}{1 - g(\vq) \ \chi_0 (\vq, \omega)} \,,
\end{align}
with the bare spin susceptibility
\begin{align}
\chi_0 (\vq, \omega) &= \dfrac{1}{\beta} \sum_{i \nu_m} \mathrm{Tr} \left[ G_{\vk,m}  \tau_0 G_{\vk + \vq,m}^+ \tau_0 \right] \notag \\
&= \iint\displaylimits_{-\infty}^{+\infty} \text{d}\omega_1 \ \text{d}\omega_2 \dfrac{ \D \DD + \omega_1 \omega_2 + \eps \epss}{2 \E \EE} \ \dfrac{n_F (\omega_1) - n_F (\omega_2)}{\omega_1 - \omega_2 + \omega + i \delta} \notag \\
&\times \left[ \delta(\omega_1 - \E) - \delta(\omega_1 + \E) \right] \left[ \delta(\omega_2 - \EE) - \delta(\omega_2 + \EE) \right] 
\end{align}
and $g(\vq) = g_0 [1 - 0.1 (\cos q_x + \cos q_y)]$.

However, we now show that the coupling of the spin channel (and therefore the spin susceptibility or the resonance peak) to either the amplitude of the superconductor or to the charge/Raman vertex is not possible in light experiments, i.e.~: in the limit of vanishing transferred momentum, $\mathbf{q} \rightarrow 0$.\\
Indeed, within this limit, we first calculate the susceptiblity which couples the spin channel ($\Psi_\vk^\dagger \tau_0 \Psi_\vk$) to the Higgs mode ($\tau_1$):
\begin{align}
\chi_{\tau_0 \tau_1} (\omega) &= \dfrac{1}{\beta} \sum_{i \nu_m} \mathrm{Tr} \left[ G_{\vk,m}  \tau_0 G_{\vk,m}^+ \tau_1 \right] \notag \\
&= \iint\displaylimits_{-\infty}^{+\infty} \text{d}\omega_1 \ \text{d}\omega_2 \dfrac{2 \D (\omega_1 + \omega_2)}{2 \E^2} \ \dfrac{n_F (\omega_1) - n_F (\omega_2)}{\omega_1 - \omega_2 + \omega + i \delta} \notag \\
&\times \left[ \delta(\omega_1 - \E) - \delta(\omega_1 + \E) \right] \left[ \delta(\omega_2 - \E) - \delta(\omega_2 + \E) \right] = 0 \,.
\end{align}
Analogously, for the coupling between the spin and the charge channel we get
\begin{align}
\chi_{\tau_0 \tau_3} (\omega) &= \dfrac{1}{\beta} \sum_{i \nu_m} \mathrm{Tr} \left[ G_{\vk,m}  \tau_0 G_{\vk,m}^+ \tau_3 \right] \notag \\
&= \iint\displaylimits_{-\infty}^{+\infty} \text{d}\omega_1 \ \text{d}\omega_2 \dfrac{2 \eps (\omega_1 + \omega_2)}{2 \E^2} \ \dfrac{n_F (\omega_1) - n_F (\omega_2)}{\omega_1 - \omega_2 + \omega + i \delta} \notag \\
&\times \left[ \delta(\omega_1 - \E) - \delta(\omega_1 + \E) \right] \left[ \delta(\omega_2 - \E) - \delta(\omega_2 + \E) \right] = 0 \,.
\end{align}
As shown by Venturini \textit{et al.} in \cite{PhysRevB.62.15204}, it is possible to obtain some contribution of the spin resonance peak even in the Raman response only including a two-magnon process. In this case, however, it has been shown that the spin response is not affecting significantly the intensity of the $A_{1g}$ Raman response, but it can rather change the peak position.

\section{VI. Experimental data}
\begin{figure}[h!]
    \centering
    \includegraphics[scale=0.15]{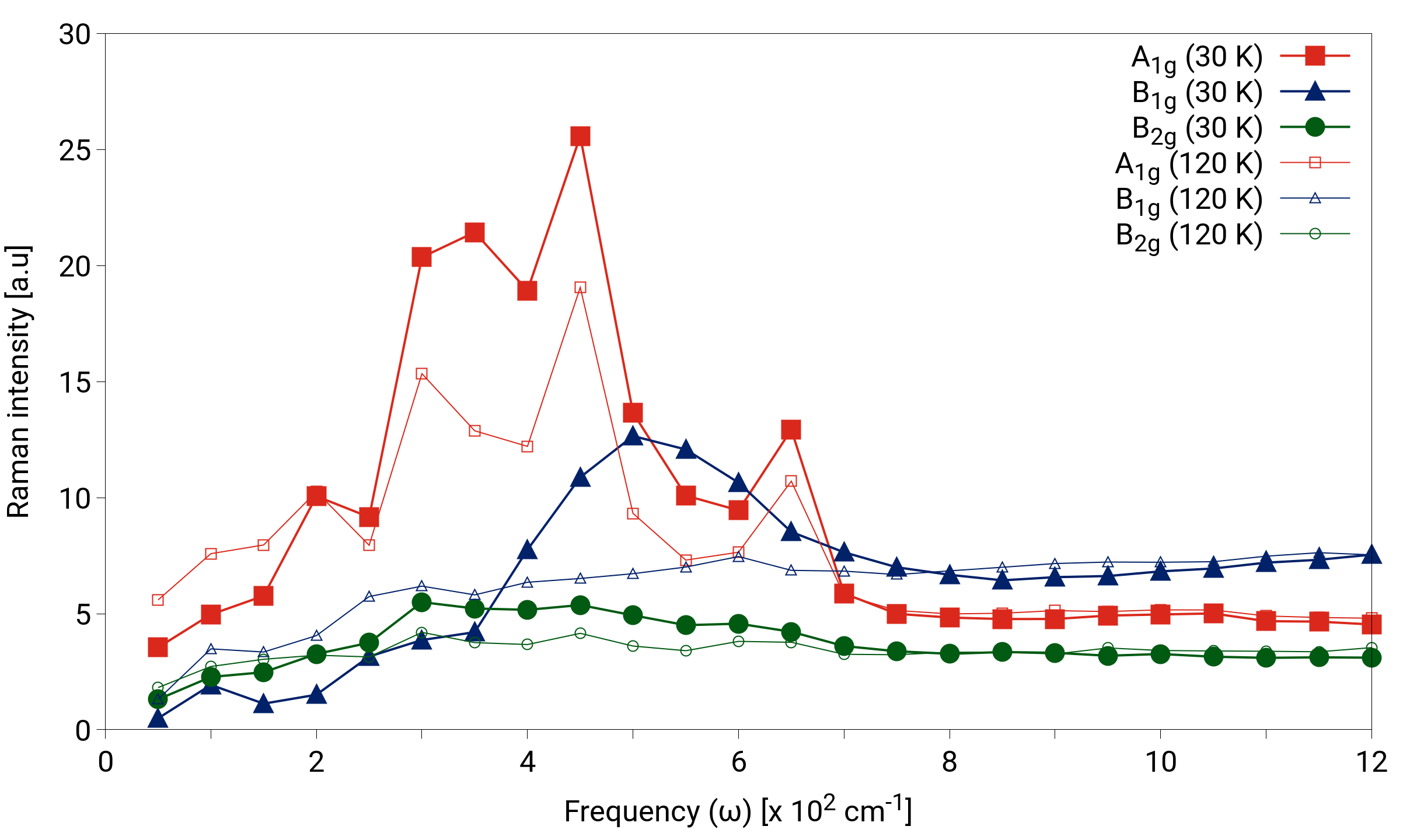}
    \caption{\textbf{Experimental raw data.} Raman spectra for different symmetries ($A_{1g}$ in red squares, blue triangles for $B_{1g}$ and green dots for $B_{2g}$) on Bi2212 sample at $T= 30$ K $< T_c$ (thick lines and full points) and $T=120$ K $> T_c$ (thin lines and empty points).}
    \label{rawdata}
\end{figure}

\begin{figure} [h!]
    \centering
    \includegraphics[scale=0.15]{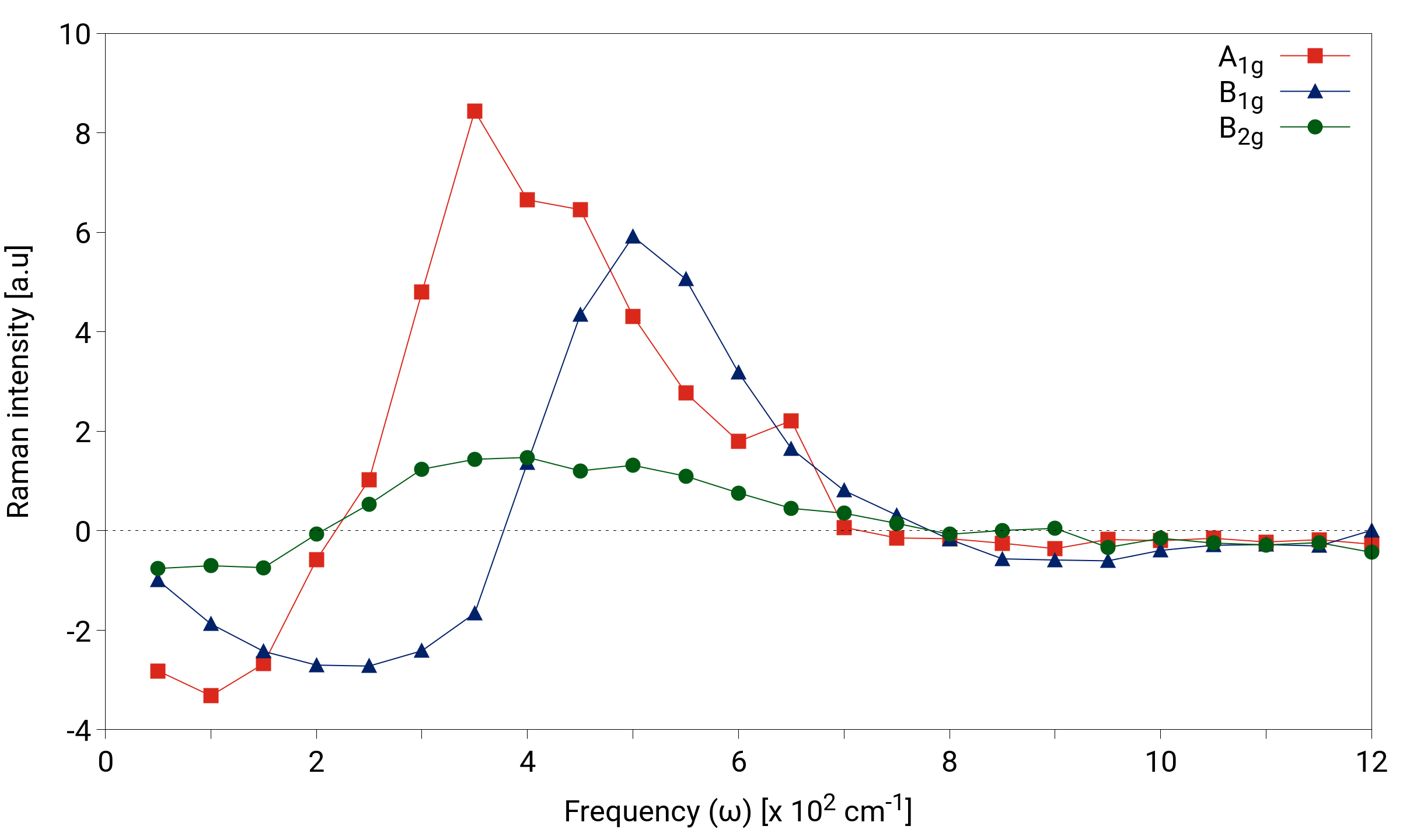}
    \caption{\textbf{Experimental superconducting Raman response}. Experimental data of the Raman spectra for different symmetries on Bi2212 sample obtained from the difference of the Raman spectra $R_{T = 30 \text{K}} - R_{T = 120 \text{K}}$ in Fig. ~\ref{rawdata}.}
    \label{exp_data}
\end{figure}

The Raman spectra were measured in high vacuum ($p<10^{-6}$ mbar) using standard equipment: they represent pure symmetry projections which were derived from the raw data by linear combinations \cite{Muschler2010}. The spectra are measured with reduced resolution to avoid the accumulation of surface layers at low temperature.  The raw data for two different temperatures, $T = 30$ K $< T_c$ and $T = 120$ K $> T_c$ are shown in Fig. ~\ref{rawdata}.  The spectra contain contributions from electronic and phononic excitations. Due to the large step width and the poor resolution of approximately 20\,cm$^{-1}$ the phonon lines are not well resolved. Only the $A_{1g}$ lines at 450 and 650\,cm$^{-1}$ are originating from phonons. In general, the phonons may be removed by subtracting the normal-state spectra from those in the superconducting state. Here, we put the plot difference $R_{T = 30 \text{K}} - R_{T = 120 \text{K}}$. 
Since the phonons depend only weakly on temperature no anomalies are seen in the spectra of Fig.~\ref{exp_data}. 
The negative intensities  in Fig.~\ref{exp_data} result from the suppression of the scattering inside the gap opening below the critical temperature$T_c=94$ K. They are suppressed in the main text.  

\newpage
\section{VII. Numerical results and parameters used}
\begin{figure}[h!]
    \centering
    \includegraphics[scale=0.23]{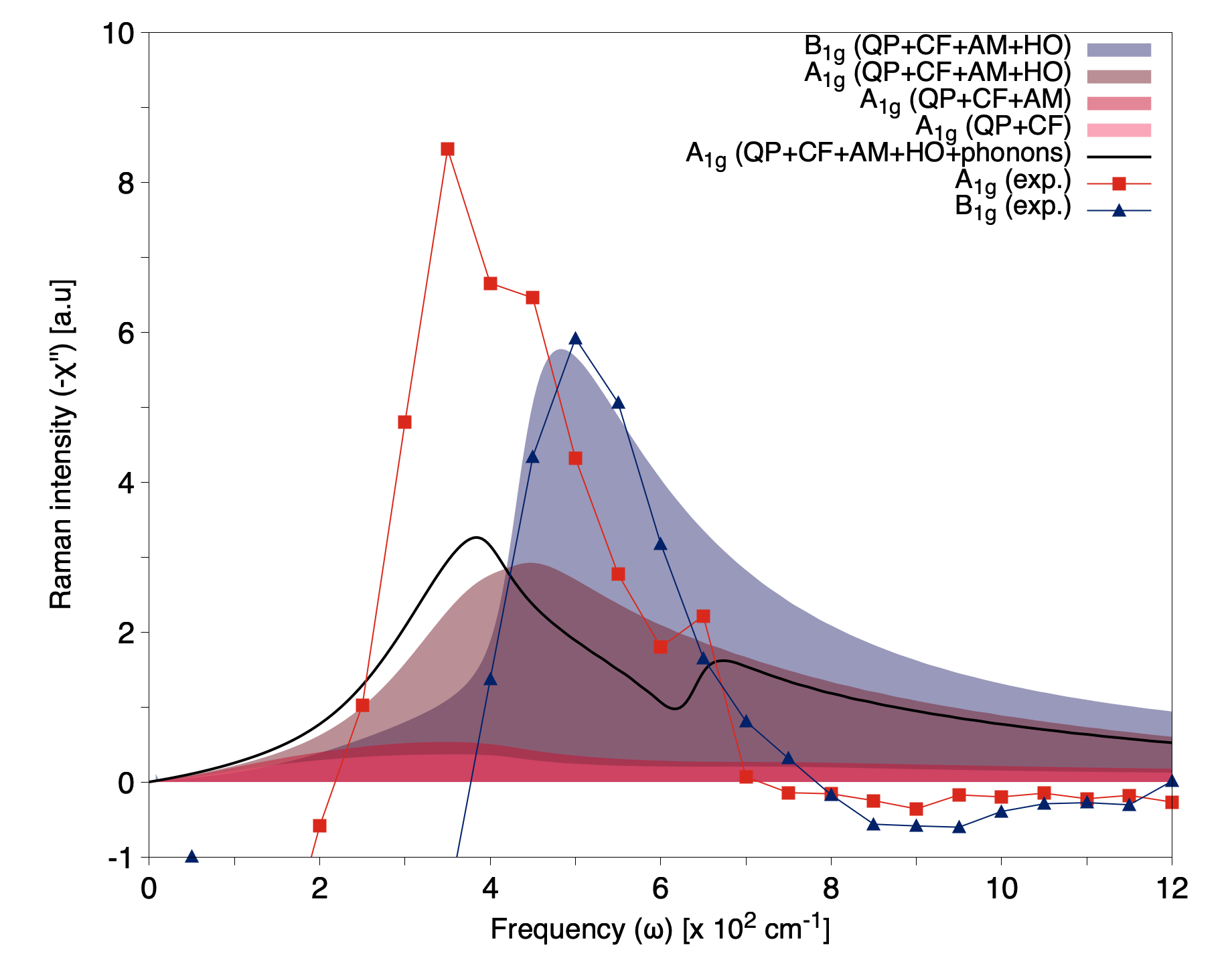}
    \caption{\textbf{Raman spectrum and phonons effects}. Comparison of Raman intensities: theoretical results (represented by filled curves)  of quasiparticles (QPs)+charge fluctuations (CFs), with amplitude mode (AM) and with Higgs oscillations (HOs) obtained with a full k-points summation over the Brillouin zone for different symmetries (blue for the $B_{1g}$ and colors in the scale of red for the $A_{1g}$), and corresponding experimental values (represented by dots, the lines connecting them are a guide for the eyes). The black line represents the $A_{1g}$ Raman response including the phonon contribution of the $A_{1g}$ phonon peaked at $\sim 80$ meV, as expressed in Eq. (\ref{phononVC}), with an electron phonon coupling strength $g_{\text{ph}}/\Delta_{\text{max}} = 0.97$. For the other parameters used: see text.}
    \label{RamanBZph}
\end{figure}

We have first calculated the Raman response with the full Brillouin zone summation of the k-points, using the parameters $\Delta_{\text{max}} = 31$ meV, broadening $\delta = 4 \ \text{meV}$, $t = 125$ meV, $B=0.65$, $\mu = -1.2$ in the band structure dispersion in Eq. (\ref{dispersion}): the result is shown in Fig. ~\ref{RamanBZph}.
In particular, considering also the effect of the $A_{1g}$ phonon peaked at $\omega_{\text{ph}} = 80$ meV (\cite{PhysRevB.45.7392}) on the Raman response, the full $A_{1g}$ Raman response (which also includes quasiparticles and many-body Higgs). The bare phonon contribution is negligible, but when the many-body Higgs interactions are included, it contributes to increase the Raman peak and to shift it towards $\omega = 350 \ \text{cm}^{-1}$, as well as to screen the response at higher frequencies revealing a shoulder at $\omega = 650 \ \text{cm}^{-1}$. These features have to be included to reach a higher qualitative agreement with the experimental data.

For the approach of the integration over the Fermi surface, as derived in Section III and whose results are shown in Fig.2 in the main text, we used the following values for the parameters which appear in Section III: $\Delta_{\text{max}} = 29$ meV, broadening $\delta = 4 \ \text{meV}$, $\alpha_1 = 136$ meV, $\alpha_1/ \beta_1 = 1.09$ and $\alpha_2 = -0.15 \ \alpha_1$. They are all reasonably within the range used in the previous literature \cite{PhysRevB.51.16336, PhysRevB.54.15547}.\\
The parameters chosen for the results in the main text correspond to those which provide the best fit of all the three symmetries $A_{1g}, B_{1g}, B_{2g}$ symultaneously.  In fact,we notice that the results in peak position and intensity of each symmetry Raman response is not independent from the others for changing the parameters involved. \\
\begin{figure}[h!]
    \centering
    \includegraphics[scale=0.13]{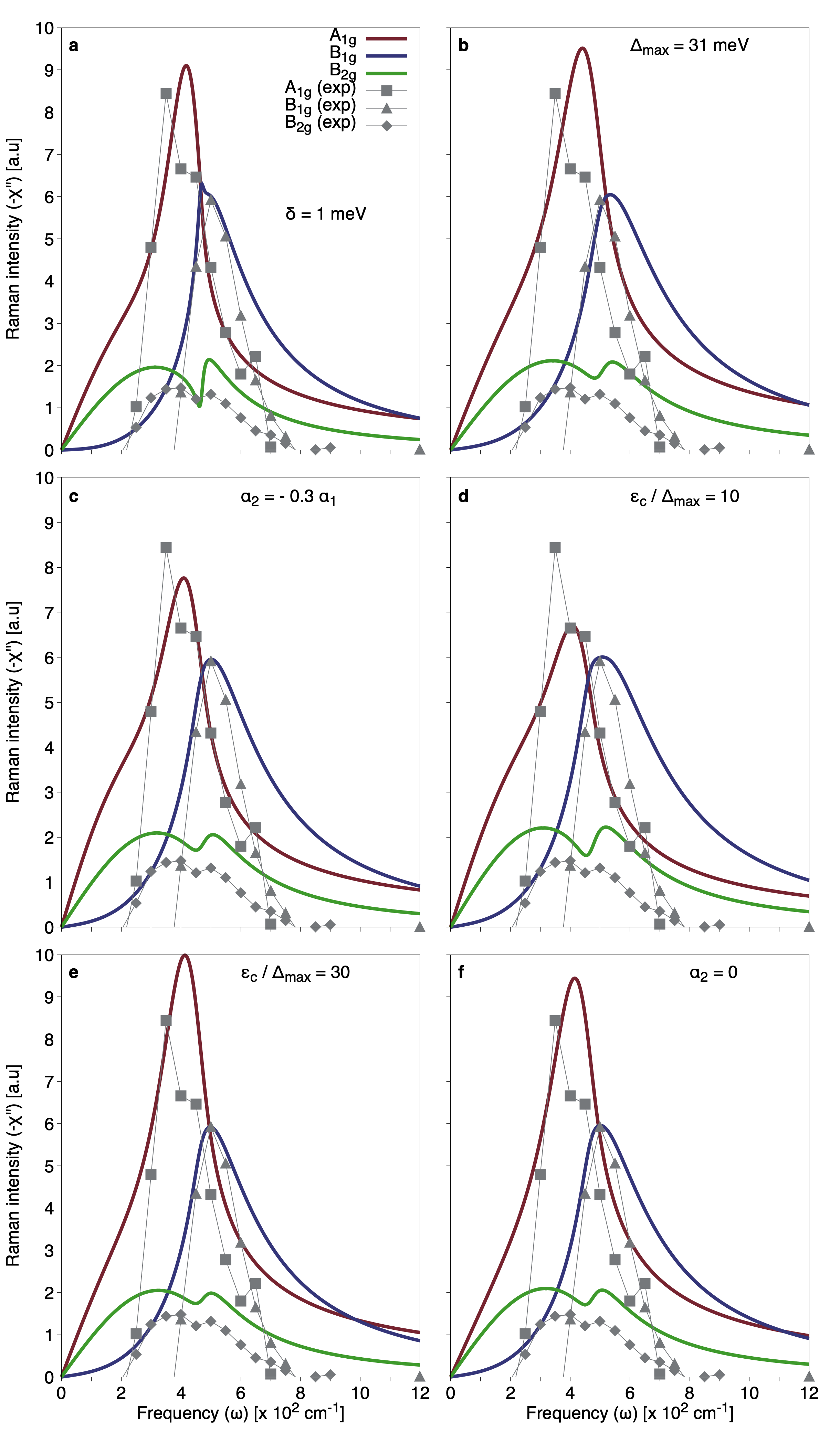}
    \caption{\textbf{Systematics Raman spectra with parameters variation}. Systematic comparison between the theoretical results (represented by filled curves: red for the $A_{1g}$, blue for the $B_{1g}$ and green for the $B_{2g}$) obtained with different parameters (see the text for the parameters used, the only varied parameter is written explicitly for each plot) which include the many-body Higgs effects, and the experimental values (represented by dots: the lines connecting them are a guide for the eyes).}
    \label{multiRaman}
\end{figure}
Therefore, we also show systematic calculations for different values of the parameters in Fig.~\ref{multiRaman}, where in each panel the parameter which differs from the optimal one is explicitly written. We notice that the actual intensities of the peak are within $10-15 \%$ with respect to the best fit, therefore showing a minor sensitivity to the parameters when changed within a reasonable interval.  In addition, we still emphasize that changing a parameter affects the three symmetries simultaneously, and the best fit is not chosen by fine tuning of the $A_{1g}$ symmetry, but on the most reasonable accordance of all the three symmetries at the same time, obtained with the same set of parameters. 
This is consistent with the robustness and the unambiguous presence of the Higgs oscillations' contribution.

\newpage
\section{VIII. Transient reflectivity change}
For linear processes, the time-dependent response $r(t)$ of a system which undergoes a time-dependent perturbation $p(t')$ is given by 
\begin{align}
    r(t) = \int_{- \infty}^{t} \text{d}t' \ f (t-t') \ p(t') \,,
\end{align}
where $f (t-t')$ is the response function. In our case the response function is the real part of the first-order susceptibility, $f (t-t') = \chi'^{(1)} (t-t')$.\\
For nonlinear processes like those occurring in terahertz pump-optical probe (TPOP) experiments, we need to use higher order susceptibilities: thus, the nonlinear polarization is given by  
\begin{align}
    P(t) = \varepsilon_0 E_{\text{probe}}(t) \int_{- \infty}^{t} \text{d}t' \chi'^{(3)}(t-t') E^2_{\text{pump}}(t') \,,
\end{align}
where the response function $\chi^{(3)}(t-t')$ is the third-order susceptibility, $P(t)$ is the response of the system (induced nonlinear polarization), while $E_{\text{pump}}$ and $E_{\text{probe}}$ are the pump and probe electrical fields, respectively. The time-dependent response function can be calculated with the inverse Fourier transform 
\begin{align}
    \chi^{(3)} (\tau = t-t') &= \mathcal{F}^{-1} \lbrace \chi^{(3)} (\nu) \rbrace (\tau) = \int_{0}^{+ \infty} \text{d}\nu \ \chi^{(3)} (\nu) \ e^{i \nu \tau} \,.
\end{align}

\begin{figure}
    \centering
    \includegraphics[width=8cm]{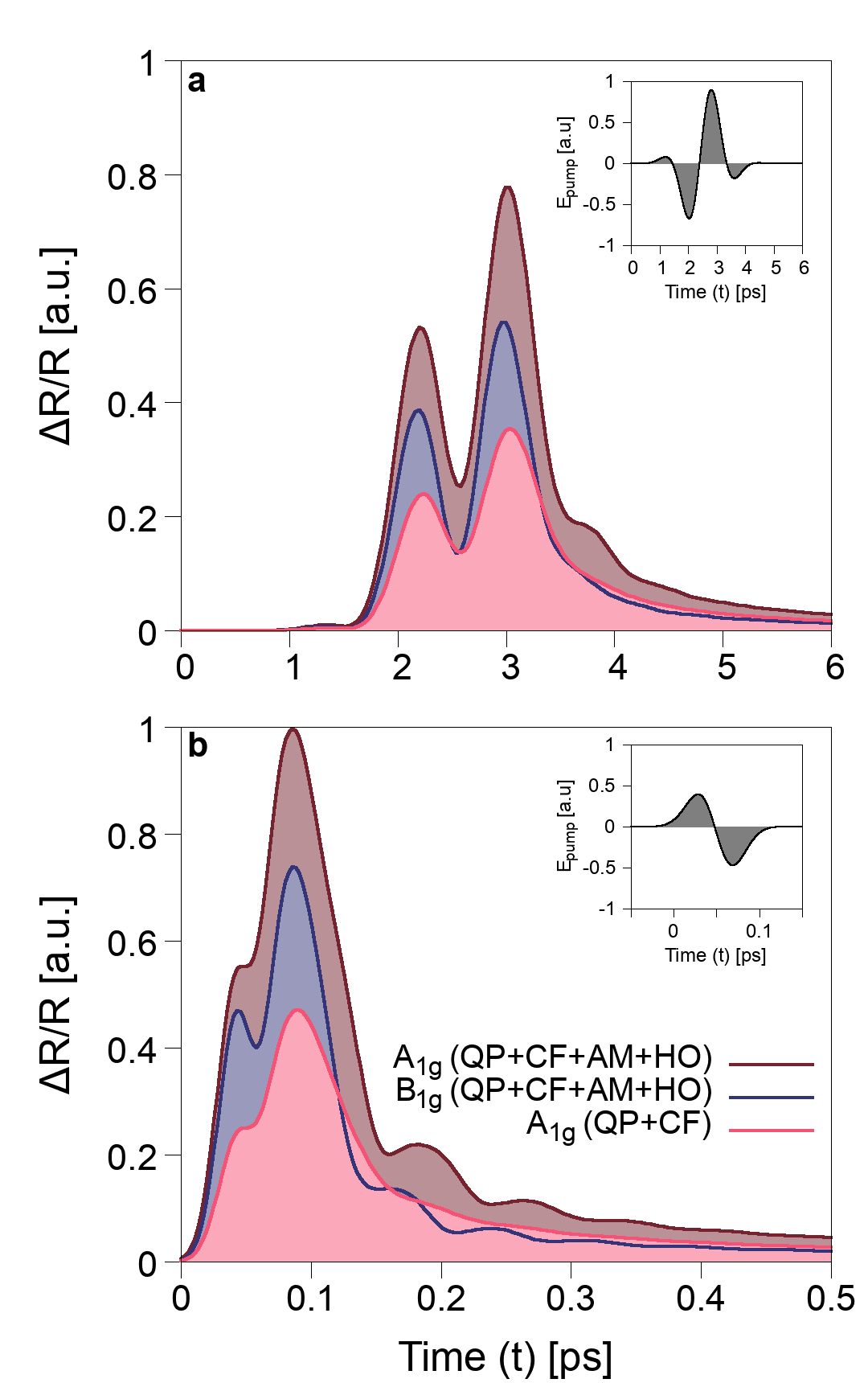}
\caption{\textbf{Transient reflectivity change}.  Normalized transient reflectivity change $\Delta \text{R} (t)/\text{R}$, calculated using the Fermi surface harmonics, for different duration ($\Delta t_{\text{pump}}$) and frequency ($\omega_{\text{pump}}$) of the electric pump pulse ($E_{\text{pump}}$, shown in the insets): $\Delta t_{\text{pump}} \sim 4 \ \text{ps}$ and $\omega_{\text{pump}} = 0.5 \ \text{THz}$ in (a), $\Delta t_{\text{pump}} \sim 0.1 \ \text{ps}$ and $\omega_{\text{pump}} = 5.57 \ \text{THz}$ in (b). The red and blue curves represent the transient reflectivity of the $A_{1g}$ and $B_{1g}$ symmetries, respectively, calculated including the full Higgs contributions. The light red curve, instead, is obtained considering only the quasiparticle contribution to the $A_{1g}$.}
    \label{DeltaR_fig}
\end{figure}

The transient reflectivity change due to a pump electric pulse $E_{\text{pump}}(t)$ is given by
\begin{align} 
    \dfrac{\Delta \text{R}}{\text{R}}(t) \propto \int_{- \infty}^{t} \chi'^{(3)}(t-t') \ E^2_{\text{pump}}(t') \ \text{d}t' \,, \label{Refl}
\end{align}
where $\chi'^{(3)} (t-t')$ is the real part of the third order susceptibility in real time, $t$ being the time at which the reflectivity change is detected, $t'<t$ the previous time at which the pump pulse interacts with the condensate.

\subsection{Calculations and results}
Recently, Katsumi et al. \cite{PhysRevLett.120.117001, PhysRevB.102.054510} demonstrated that the relative intensity between the transient reflectivity change projected onto the symmetries $A_{1g}$ and the $B_{1g}$ in a TPOP experiment on Bi2212 cannot be explained only with the QP contribution, thus deducing that the Higgs should play an important role.\\
Here we considered an adiabatic electric pulse (in analogy to \cite{PhysRevLett.120.117001}) with the form
\begin{align}
    E_{\text{pump}} (t) = \sin{(2 \pi \omega_p t + \varphi)} \ e^{-(t-t_0)/(2 \sigma^2)} \,,
\end{align}
with frequency $\omega_p = 0.53$ THz, $\varphi = 1.65$ rad, $t_0 = 2.55$ ps, $\sigma = 0.63$ ps. Then, we used Eq. (\ref{Refl}) to evaluate the time-dependent transient reflectivity change on a $d$-wave superconductor, inserting the third-order susceptibility for a fixed symmetry containing the full many-body Higgs response (in analogy with the Raman calculation of the previous sections). 
The result, plotted in Fig. ~\ref{DeltaR_fig} \textbf{a}, clearly explains the maximum value of $\Delta \text{R}/ \text{R}$ in the $A_{1g}$ symmetry higher than the $B_{1g}$, despite the hypothesis raised in Katsumi's paper of the necessity to consider also strong correlation effects, such as the retarded phonon-mediated interaction in the $s$-wave case. \\
Moreover, we predict the appearance of characteristic oscillations when a shorter THz pump ($\omega_p = 5.57$ THz, $\sigma = 0.022$ ps, see the inset of Fig. ~\ref{DeltaR_fig}\textbf{b}) is used, simulating a quench experiment \cite{PhysRevB.101.184519}. As shown in Fig. ~\ref{DeltaR_fig}\textbf{b}, including the full many-body Higgs contribution to the third-order susceptibility, in this non-adiabatic regime oscillations with a period $T \sim 0.08 \ \text{ps}$ appear in the decaying region of $\Delta \text{R} (t)/ \text{R}$, while a lower intensity and no oscillations are present considering only the quasiparticles and charge fluctuations contributions to the $A_{1g}$ response.

\bibliography{Bibliography.bib}